%% file: 4FHL.tex
\documentclass[trackchanges,twocolumn,resetfootnote]{aastex7}

\usepackage{subcaption}
\usepackage{xcolor}
\usepackage{bm}
\begin{document}

\title{4FHL: The Fourth Catalog of Hard {\it Fermi}-LAT Sources}

\input{authors_orcid}

\collaboration{all}{The {\it Fermi}-LAT Collaboration}

\begin{abstract}

We present a catalog of sources detected above 50 GeV by the {\it Fermi} Large Area Telescope (LAT) on board the {\it Fermi} Gamma-ray Space Telescope using 16 years of observations. This is the Fourth Catalog of Hard {\it Fermi}-LAT Sources (4FHL), containing 673 objects detected in the 50 GeV-2 TeV energy range. The sensitivity and localization are improved by factors of $\sim$2 relative to the previous LAT catalog at the same energies (i.e. the 2FHL catalog). The increased photon statistics, nearly doubling the number of detections relative to 2FHL, enables more detailed population studies, including the identification of fainter sources detected above 50 GeV. The  majority of the sources (84.7\%) are associated with extragalactic counterparts, primarily BL Lacertae objects, including 36 objects at redshift $z>1$. Galactic sources account for 10\% of the sample, mainly pulsar wind nebulae and supernova remnants, while $\sim$5\% remain unassociated or of uncertain classification. Twenty sources are reported with no counterpart in previous LAT catalogs. The 4FHL catalog provides the deepest all-sky survey above 50 GeV to date. By focusing on this high-energy regime, it provides a fundamental link between space-based gamma-ray observations and the TeV range probed by current and future ground-based observatories. It serves as a useful reference for guiding future observations, including new potential targets for follow-up with Cherenkov telescopes.

\end{abstract}


\setcounter{footnote}{0}
\renewcommand{\thefootnote}{\arabic{footnote}}

\section{Introduction}
The Large Area Telescope \citep[LAT;][]{Atwood_2009} on board the \textit{Fermi} Gamma-ray Space Telescope has been continuously surveying the entire sky since its launch in August 2008. Over the course of 16 years of observations, the LAT has made significant contributions to studies of the high-energy sky. The most recent broadband all-sky LAT catalog, 4FGL-DR4 \citep{4fgl-dr4}, constructed from 14 years of data, includes 7,194 sources, providing an in-depth characterization of the gamma-ray sky from 50 MeV to 1 TeV. Typically, \textit{Fermi}-LAT catalogs are representative of the GeV sky rather than the sub-MeV or TeV sky, as the instrument’s sensitivity is optimized for the low-GeV energy band.
However, the detection of faint and hard spectrum sources is often better enhanced through dedicated higher-energy analyses.

To explore the {\it Fermi} higher-energy regime, the LAT Collaboration has released a series of hard-source catalogs optimized for energies above tens of GeV. The First {\it  Fermi}-LAT Hard Source Catalog (1FHL; \citealt{1FHL}) reported 514 sources detected above 10\,GeV,  up to 500\,GeV in the first 3 years of data. This was followed by the Second Hard Source Catalog (2FHL; \citealt{2FHL}), which used 6.7 years of LAT data and was the first to incorporate the Pass~8 event reconstruction \citep{atwood2013pass8realizationfermilat}. The 2FHL catalog, which presented 360 sources, focused on the 50\,GeV$-$2\,TeV range, where the LAT benefited from improved angular resolution and reduced diffuse background, making it particularly effective despite limited photon statistics.  The Third Catalog (3FHL; \citealt{3FHL}) extended the 1FHL analysis using 7 years of Pass~8 data, detecting 1,556 sources in the 10\,GeV$-$2\,TeV range, benefiting from enhanced background rejection and the implementation of a point spread function (PSF)-type event classification.

In this paper, we present the Fourth Catalog of Hard {\it Fermi}-LAT Sources (4FHL), which extends the work of 2FHL using 16 years of Pass 8 data \citep{bruel2018fermilatimprovedpass8event}. 

Above 50 GeV, the sensitivity improves faster than in the background-limited regime at lower energies. The lower photon statistics in this regime also make an unbinned likelihood analysis optimal, improving sensitivity to faint sources and source localization. Assuming a standard Euclidean logN-logS distribution, a factor of two increase in high-latitude sources relative to the 2FHL catalog is expected. As a result, the new catalog includes 673 sources, significantly expanding the population of LAT-detected sources above 50 GeV. 

It offers insights into extreme particle acceleration environments, such as pulsar wind nebulae  \citep[PWNe, e.g.][]{Acero2017}, supernova remnants  \citep[SNRs e.g.][]{SNRs}, and blazars \citep[e.g.][]{paliya19,denBerg2019}, while also supporting multimessenger studies \citep[e.g.][]{IceCubemulti} by linking high-energy $\gamma$ rays with other astrophysical phenomena, which might be sources of neutrinos and cosmic rays \citep[e.g. Galactic PeV accelerators,][]{marchesi20242fhlj174513035newlydiscoveredpowerful}, remaining relevant in broader contexts, including cosmology \citep[e.g.][]{dominguez24,EBL_Greaux_2024,abdalla24,furniss25} and dark matter \citep[e.g.][]{charles16,Coronado_Bl_zquez_2019,buehler20}.


With its all-sky coverage and long-term exposure over an energy range extending up to 2 TeV, the 4FHL catalog complements ground-based Cherenkov telescopes such as VERITAS \citep{VERITAS}, MAGIC \citep{MAGIC}, and H.E.S.S. \citep{HESS} by providing a well-characterized high-energy source sample that can guide targeted TeV observations. It serves as a reference for multiwavelength follow-up studies and supports the identification and study of extreme particle accelerators through source selection, follow-up prioritization, population studies, and comparisons with recent and future observatories, including HAWC \citep{3HWC}, LHAASO \citep{LHAASO}, SWGO \citep{SWGO}, and CTAO \citep{CTAO_2,CTAO}.

The paper is organized as follows. In \S~\ref{sec:analysis}, we describe the data selection, source detection, and association procedure used to construct the catalog. 
In \S~\ref{sec:4FHL}, we present the 4FHL catalog main results, including a description of its content and the general properties of the detected sources. Finally, we summarize the main results in \S~\ref{sec:summary}.

 
\section{Analysis}\label{sec:analysis}

\subsection{Data Selection}\label{sec:data_select}
\begin{figure}[ht]
    \centering
    \includegraphics[width=1\columnwidth]{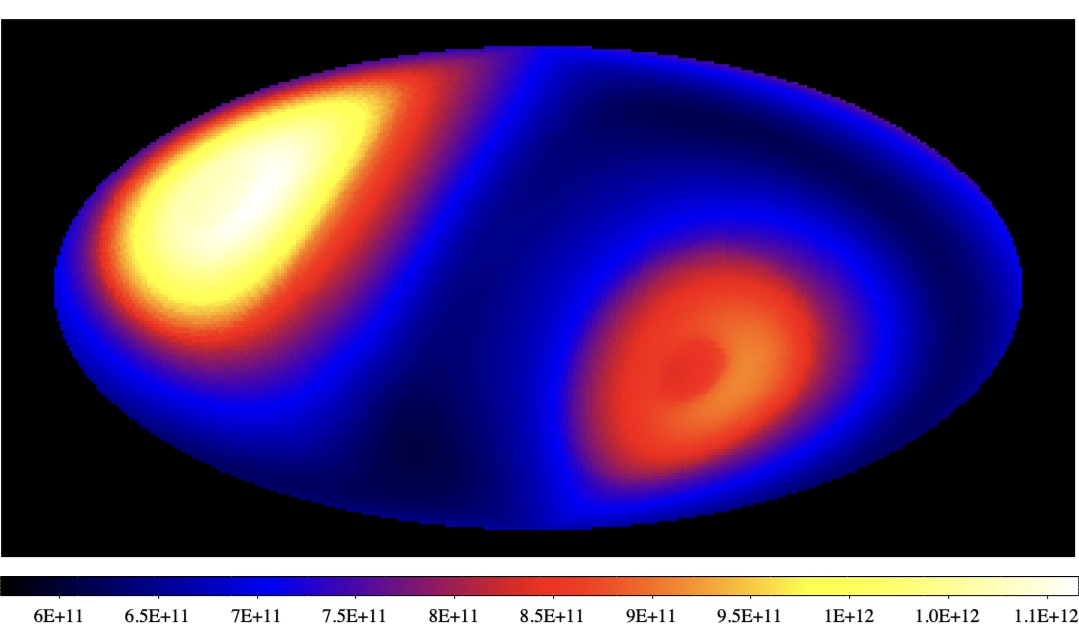}
    \caption{Exposure map at 50 GeV  in Galactic coordinates and Hammer-Aitoff projection for the 16 year 
    period, with exposure values in $\text{cm}^2 \, \text{s}$. The median exposure is $1.11 \times 10^{12} \, \text{cm}^2 \, \text{s}$. The maximum exposure is reached at the North Celestial Pole, while the minimum occurs along the celestial equator.}

    \label{fig:expmap}
\end{figure}

We used 16 years of \textit{Fermi}-LAT data, covering the period from 2008 August 4 to 2024 August 30 (Mission Elapsed Time 239,557,417 to 746,711,954.5 s). We selected Pass~8 Release~3 (P8R3, \citeauthor{bruel2018fermilatimprovedpass8event}~\citeyear{bruel2018fermilatimprovedpass8event}) SOURCE class events with reconstructed energies in the 50\,GeV to 2\,TeV range. To reduce contamination from Earth's limb emission, photons with zenith angles greater than $105^\circ$ were excluded. In addition, we removed time intervals during which the LAT was not in nominal science operations, requiring that the data quality and configuration flags satisfy \texttt{DATA\_QUAL > 0} and \texttt{LAT\_CONFIG == 1}. No rocking angle cut was applied, as its impact is minimal above 50\,GeV.
The exposure map shown in Figure \ref{fig:expmap} reveals the variation in the LAT exposure across the sky at 50\,GeV. The median exposure value is $1.11 \times 10^{12} \text{cm}^2 \text{s}$ for the 16 year period, with deviations from this median ranging from $-48\%$ to $+70\%$. These variations are primarily due to differences in declination. The exposure at southern declinations is reduced due to the interruption of LAT data taking during South Atlantic Anomaly passages. The exposure reaches its maximum near the north celestial pole, while the minimum occurs along the celestial equator. A region near the southern celestial pole shows a slight decrease in exposure due to the zenith angle limitation of $105^\circ$.

\begin{figure}[ht]
    \centering
    \includegraphics[width=1\columnwidth]{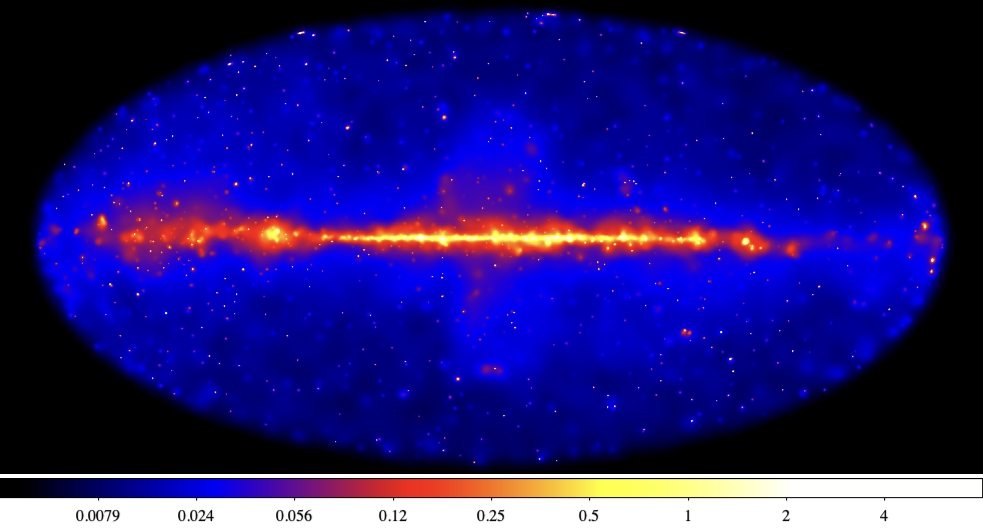}
    \caption{Adaptively smoothed 2D count map in the 50 GeV-2 TeV band represented in Galactic coordinates and Hammer-Aitoff projection. The image has been smoothed with a Gaussian kernel whose size was varied to achieve a minimum signal-to-noise ratio under the kernel of 2. The color scale is logarithmic, and the units are counts per (0.1$^\circ$)$^2$.}
    \label{fig:skymap}
\end{figure}

The final photon selection includes approximately 133,400 events across the full sky, corresponding to about 3 photons per square degree. The resulting smoothed sky count map at $>$50 GeV, shown in Figure~\ref{fig:skymap}, reveals both point-like sources and large-scale diffuse structures, particularly along the Galactic plane and in the regions associated with the \textit{Fermi} bubbles \citep{Su_2010,Ackermann_2014}.

\subsection{Source Detection}\label{sec:source_detect}
The first step in source detection involves identifying potential point-like seeds. Seeds are identified via (1) a sliding-cell algorithm as excesses above the background and (2) via a wavelet analysis using the \texttt{PGWave} tool \footnote{The sliding-cell step used \texttt{batcelldetect} (\url{https://swift.gsfc.nasa.gov/analysis/threads/batsrcdetect.html}; see also \citealt{cell-sliding}) with a 0.1$^{\circ}$ PSF and S/N = 4.5, while \texttt{PGWave} was run conservatively with 0.1$^{\circ}$ pixels, a 0.23$^{\circ}$ scale, and a 5$\sigma$ threshold.} \citep{pgwave_2}, also employed in the 1FLE catalog \citep{1FLE}.
Additionally, sources from the 3HSP \citep{3HSP} catalog were incorporated to enhance the analysis of the high-frequency peaked BL Lac population, as they are the dominant source class above 50 GeV. We also included 4FGL-DR4 seed candidates identified with \texttt{pointlike} \citep{kerr2010}. This procedure starts from the 4FGL-DR4 source list, reoptimizing their positions and power-law spectral parameters over 16 years of LAT data, identifying new candidates from peaks in the residual TS maps. Newly identified seeds were added to the model and the process was iterated. For further details on the \texttt{pointlike} detection procedure, see \S 3.1 of the 4FGL catalog \citep{4FGL}.

For extended sources, we adopted the spatial templates and parameters reported in previous LAT catalogs, without performing a new extension analysis or a dedicated search for additional extended sources. We included the 35 sources from the 2FGES catalog \citep{2FGES} (excluding those classified as dubious), which reports extended sources detected above 10 GeV in the inner Galactic plane. For these sources, we used the azimuthally symmetric 2D Gaussian spatial models and parameters adopted in 2FGES. In addition, all extended sources from the 4FGL-DR4 catalog were considered, excluding any source whose spatial extension overlaps by more than $0.2^{\circ}$ with any 2FGES object, using the spatial model parameters and templates available in the 4FGL-DR4 source catalog\footnote{\url{https://fermi.gsfc.nasa.gov/ssc/data/access/lat/14yr_catalog/}}. The final result was an initial list of over 11,000 seeds, covering the entire sky.

To test these seeds, we performed an unbinned maximum likelihood\footnote{Unbinned analysis tutorial available at \url{https://fermi.gsfc.nasa.gov/ssc/data/analysis/scitools/likelihood_tutorial.html}} (ML) analysis. Typically, LAT data are reduced using a binned ML algorithm, where data are divided into discrete bins. However, some information can be lost in the binning process, making it less sensitive to faint sources or subtle features. In contrast, the unbinned ML analyzes each event individually, considering its energy and position. This approach improves sensitivity to faint sources and source localization; however, it does not account for energy dispersion, which may affect the results near 1 TeV. Although unbinned ML analysis is computationally intensive, it is optimal for the high-energy regime above 50 GeV, where photon statistics are lower. The general methodology followed the pipeline used in the 2FHL catalog, which was carefully developed and tested. The specific steps for the analysis are outlined below.

The analysis was conducted across 154 regions of interest (ROIs), with radii ranging from 5° to 20°, whose sizes and positions were optimized so that every seed falls within an ROI, while ensuring that no region contains more than 45 seeds allowed to vary. This keeps the number of free parameters per ROI $<100$, which is within the regime of reliable MINUIT convergence. For each ROI, a sky model was created that includes all potential seeds in the region, along with the Galactic and isotropic diffuse emission models. We used the latest background models\footnote{Available at \url{https://fermi.gsfc.nasa.gov/ssc/data/access/lat/BackgroundModels.html}}, \texttt{gll\_iem\_v07.fits} for the Galactic diffuse emission, and \texttt{iso\_P8R3\_SOURCE\_V3\_v1.txt} for the isotropic component, extrapolated up to 2 TeV using a power law with the same photon index as above 10 GeV ($\Gamma \sim 1.9$). We considered the \texttt{P8R3\_SOURCE\_V3} LAT instrument response functions (IRFs).  The adopted diffuse models and IRFs correspond to the recommended LAT configuration for the P8R3 SOURCE event class used in this analysis. Above 50 GeV, the narrow LAT PSF and the photon-limited regime reduce the impact of diffuse components, making the analysis less sensitive to the background.

The fitting procedure was conducted iteratively to ensure convergence and achieve an optimal solution. Multiple \texttt{FermiTools} 2.2.0\footnote{For more information on FermiTools, refer to \url{https://fermi.gsfc.nasa.gov/ssc/data/analysis/software/}} were used according to the specific requirements of each stage.
The 4FHL catalog was constructed using the following pipeline:

\begin{enumerate}
\item Complex ML fits require approximate starting values for the model parameters. Therefore, the first step consists of fitting each source separately with a power-law\footnote{Given the limited photon statistics in the 50 GeV-2 TeV range, no statistically significant spectral curvature, including the curvature induced by EBL attenuation in extragalactic sources, is expected to be detectable for individual sources by the LAT within the 50 GeV-2 TeV energy band.} spectral model to determine its initial spectral parameters. During this process, the parameters of the diffuse emission models were allowed to vary freely. The significance of each source was assessed using the test statistic ${\rm TS}=2(\ln \mathcal{L}_1 - \ln \mathcal{L}_0)$, where $\mathcal{L}_0$ and $\mathcal{L}_1$ are the likelihoods of the background (null hypothesis) and the source-plus-background hypothesis, respectively. Sources with ${\rm TS} < 10$ were removed from the model at each step. Once the spectral parameters and significance were determined, a global fit was performed, allowing all parameters of sources with ${\rm TS} \geq 10$ to vary. A second global fit was performed after removing sources with ${\rm TS} < 10$ from the previous global fit. Throughout this step, as well as the rest of the pipeline, spatially extended seeds were included.

\item In the second step, the positions of point-like sources were optimized using the best-fit model derived in step 1, with the {\tt gtfindsrc} tool. The optimization was performed iteratively, starting with the most significant sources and then refining the positions of fainter sources.

\item The parameters and significances of the sources were re-estimated (as in step 1) using the best-fit positions obtained in step 2. This step produces the final best-fit sky model for each ROI. Seeds with $10 \leq {\rm TS} < 25$ (445 seeds), corresponding to detections below about $4\sigma$ for a source model with four free parameters, were included in the model but are not reported in the final catalog. Following the approach used in previous LAT source-search analyses \citep[e.g.,][]{2FHL}, retaining these sub-threshold seeds provides a more flexible description of the background, accounting for possible residual mismodeled emission and reducing the risk of attributing it to nearby cataloged sources.

\item For each source, we estimated the energy of the highest-energy photon (HEP) that the fit attributes to the source model. This was done using the tool {\tt gtsrcprob}, selecting the HEP with a probability greater than 85\% to belong to the source. 

\item A spectrum with three logarithmically spaced bins (50\,GeV, 171\,GeV, 585\,GeV, 2\,TeV) was generated for each source in the ROI detected with ${\rm TS} \geq 25$, and the predicted number of source photons (N$_{\rm pred}$, reported by the likelihood) of at least three, ensuring a minimum level of photon statistics for each detection. 

\end{enumerate}

The detection procedure, adopting the same selection criteria as 2FHL, yielded 673 sources with ${\rm TS} \geq 25$ and ${\rm N}_{\rm pred} \geq 3$.  To estimate the fraction of spurious detections, we followed \citet{2fhlloglogs} and used \texttt{gtobssim} simulations including only the diffuse components, generated from 30 GeV (to account for energy dispersion), with the same pointing and livetime histories and event selection as in the real analysis. The simulated events were analyzed in the same way as the real data (as detailed above), resulting in six spurious sources (none within $1^\circ$ of a 4FHL source)  corresponding to a false-positive fraction of $<1\%$ in the final catalog.

\subsection{Source association}\label{sec:association}
The association procedure follows the same approach as that employed for the FGL catalogs and makes use of two methods, the Bayesian method and the Likelihood-ratio (LR) method \citep{1FGL,2LAC}.  High-confidence associations (probability $>0.8$) are found for 642 sources, the Bayesian and LR methods providing 628 and 569 of them, respectively, with 555 in common. The list of counterpart catalogs used in the Bayesian method is the same as for the 4FGL-DR4 release, with the addition of the 3HSP catalog \citep{3HSP} of blazars. 
Catalogs have been updated to the latest available versions whenever possible, most relevant for 4FHL being TeVCat \footnote{\url{http://tevcat.uchicago.edu/}} and the radio fundamental catalog \citep{RFC} of blazars.
The LR method is the same as described  in the previous LAT catalogs, with the addition of eROSITA Data Release 1 (eRASS1) for the X-band associations.

A total of 610, 525, and 300 4FHL sources were already included in 4FGL-DR4, 3FHL, and 2FHL, respectively, while 20 have no nearby counterpart in previous LAT catalogs (see \S \ref{sec:new_gal} and \S\ref{sec:new_high-lat}). Table \ref{tab:classes} summarizes the association results. Here we distinguish between \emph{associated} and \emph{identified} sources: associations rely primarily on positional coincidence, whereas identifications require additional evidence such as correlated multiwavelength variability or resolved spatial extension.

In order to assess the systematic uncertainty in the LAT source positions, we used the positions of highest-confidence (probability $>$90\%) counterparts to calibrate the 68\% containment radius ($R_{68}$), relative to the statistical containment radius reported by \texttt{gtfindsrc} ($R_{68,\mathrm{stat}}$), following the method outlined in \cite{3FGL}.

The correction is parametrized by two factors: $f_{rel}$, which adjusts $R_{68,\mathrm{stat}}$ as a percentage correction, and $\Delta_{abs}$, which corrects for a systematic offset:


\begin{equation}
 R_{68} = \sqrt{(R_{68,\text{stat}} \times f_{rel})^2 + \Delta_{abs}^2}
\end{equation}

We assume a 2D Gaussian distribution for the angular separation between a 4FHL source and its corresponding high-confidence counterpart. Taking $\Theta$ as the angular separation, we define


\begin{equation}
 y = \frac{1}{2} \left(1.51 \frac{\Theta}{R_{68}}\right)^2,
\end{equation}

where $R_{68}$ is 1.51 $\sigma$ of the 2D Gaussian distribution. $e^{-y}$ then follows a uniform distribution between 0 and 1. We determine the correction parameters $f_{rel}$ and $\Delta_{abs}$ by minimizing the $\chi^2$ between the histogram of observed $e^{-y}$ values and the uniform distribution.
The best-fit values for the correction factors were found to be \(f_{rel} = 1.09\) and \(\Delta_{abs} = 0.004^\circ\). Similar values were used in the 2FHL analysis (\(f_{rel} = 1.08\), \(\Delta_{abs} = 0.003^\circ\)).

\section{The 4FHL catalog}\label{sec:4FHL}

The 4FHL catalog contains a total of 673 sources detected across the entire sky, including 55 LAT extended sources. All  sources satisfy the detection criteria of ${\rm TS} \geq 25$ and N$_{\rm pred} \geq 3$. Compared to the 2FHL catalog, which was based on 6.7 years of LAT data and reported 360 sources, the 4FHL nearly doubles the number of detections. This increase reflects the steady improvement in sensitivity above 50 GeV, which scales approximately linearly with observing time.

\begin{figure}[htpb]
    \centering
    \includegraphics[width=1\columnwidth]{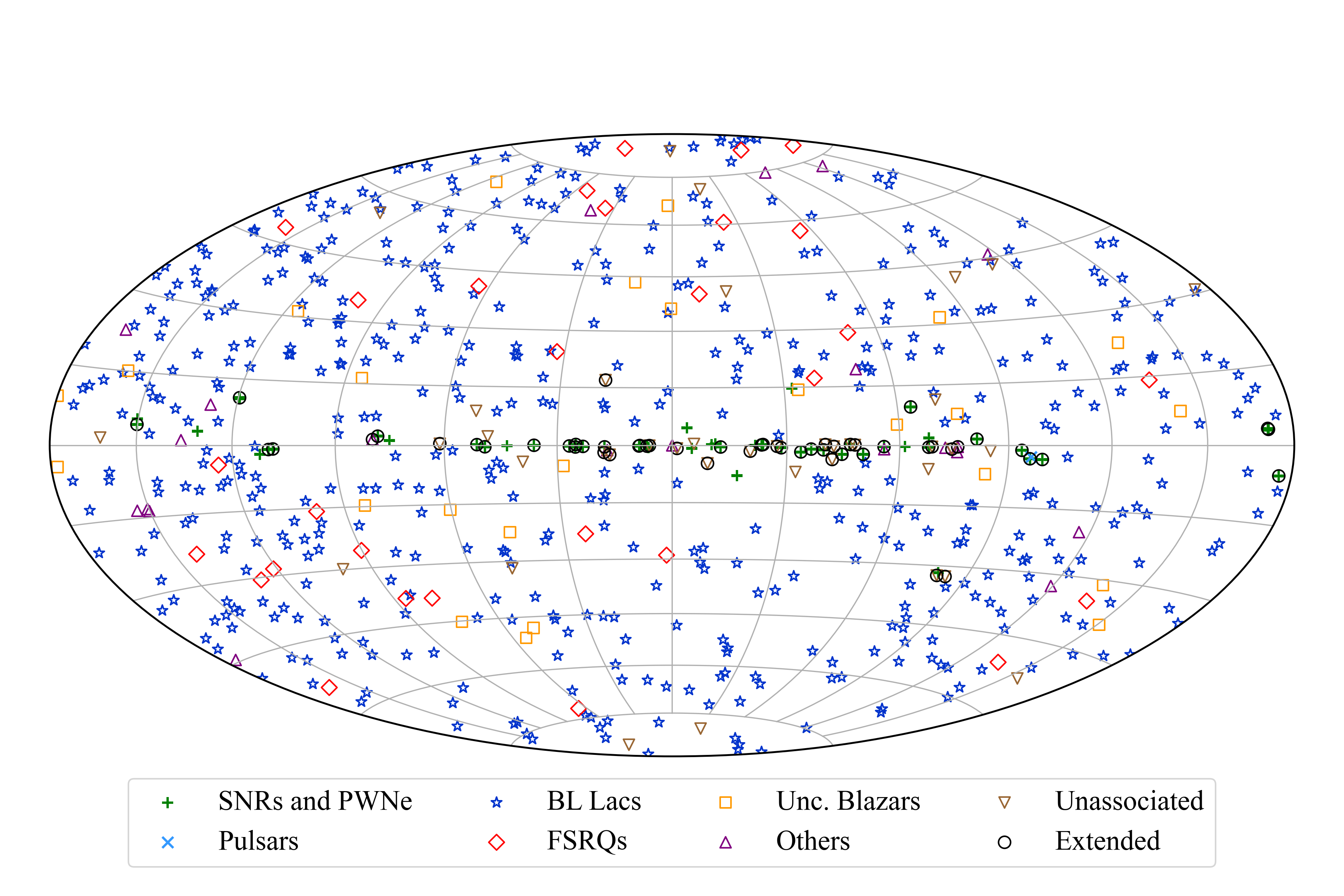}
    \caption{ Sky map in Galactic coordinates of the 4FHL sources shown in a Hammer--Aitoff projection, classified according to their most likely association class.}
    \label{fig:spatial_distribution}
\end{figure}

The association procedure (see \S~\ref{sec:association}) found that the 4FHL population is predominantly extragalactic (570 sources), representing 84.5\% of the detections, while 10\% (65 sources) are associated with Galactic sources. The remaining $\sim 5$\% (38 sources) are classified as unassociated (31) or associated with sources of unknown nature\footnote{\label{fn:unk}Unknown sources are 4FHL sources associated with  X-ray or radio counterparts whose nature is not known.} (7).  Among the 31 sources that remain unassociated, 16 of them are extended.  Out of the 15 others, seven lie within less than 10\arcmin\  of a gamma-ray blazar detected by the LAT at lower energy (i.e. 4FGL), most often (5 out of 7) of the BL Lac type.
Moreover, 63 sources were not included in the 4FGL-DR4 catalog, and 20 sources do not have an associated counterpart previously reported in any {\it Fermi} catalog or TeVCat before. These sources are discussed in more detail in \S~\ref{sec:new_gal} and \S~\ref{sec:new_high-lat}.
Figure~\ref{fig:spatial_distribution} shows the spatial distribution and classification of sources in the 4FHL catalog color-coded by class. 
\subsection{Description of the Catalog}

The FITS format of the 4FHL catalog follows the standard structure adopted in previous {\it Fermi}-LAT catalogs. A detailed description of the column contents is provided in Table~\ref{tab:description4fhl}. The file includes three binary table extensions: the main catalog, the list of extended sources, and information on the ROIs.

Table~\ref{tab:description4fhl} describes the 36 columns of the main catalog. It lists the 4FHL name, source position (including Equatorial J2000 and Galactic coordinates), positional uncertainties (at 68\% and 95\% confidence), and detection significance above 50 GeV. Spectral properties are derived assuming a power-law model and are summarized by the observed photon index and its statistical $1\sigma$ uncertainty. Integrated photon and energy fluxes are provided for the full 50\,GeV$-$2\,TeV band, as well as for three logarithmically spaced sub-bands (50$-$171\,GeV, 171$-$585\,GeV, and 585\,GeV$-$2\,TeV), with the corresponding $1\sigma$ uncertainties and significances. Additional information includes the number of predicted photons (N$_{\rm pred}$) and the HEP detected, together with its probability of association to the source. The catalog also provides the most likely source association (with probability $>80\%$) and, when available, its redshift. Corresponding names from previous LAT catalogs (e.g. 2FHL, 3FHL and 4FGL) are reported, together with the most likely association at TeV energies if any. 

\input{tables/Table_class}

\input{tables/Table_description}

\subsection{General Properties }

\begin{figure*}[htbp]
\vspace{-1cm}
    \centering
    
    \includegraphics[width=0.6\textwidth]{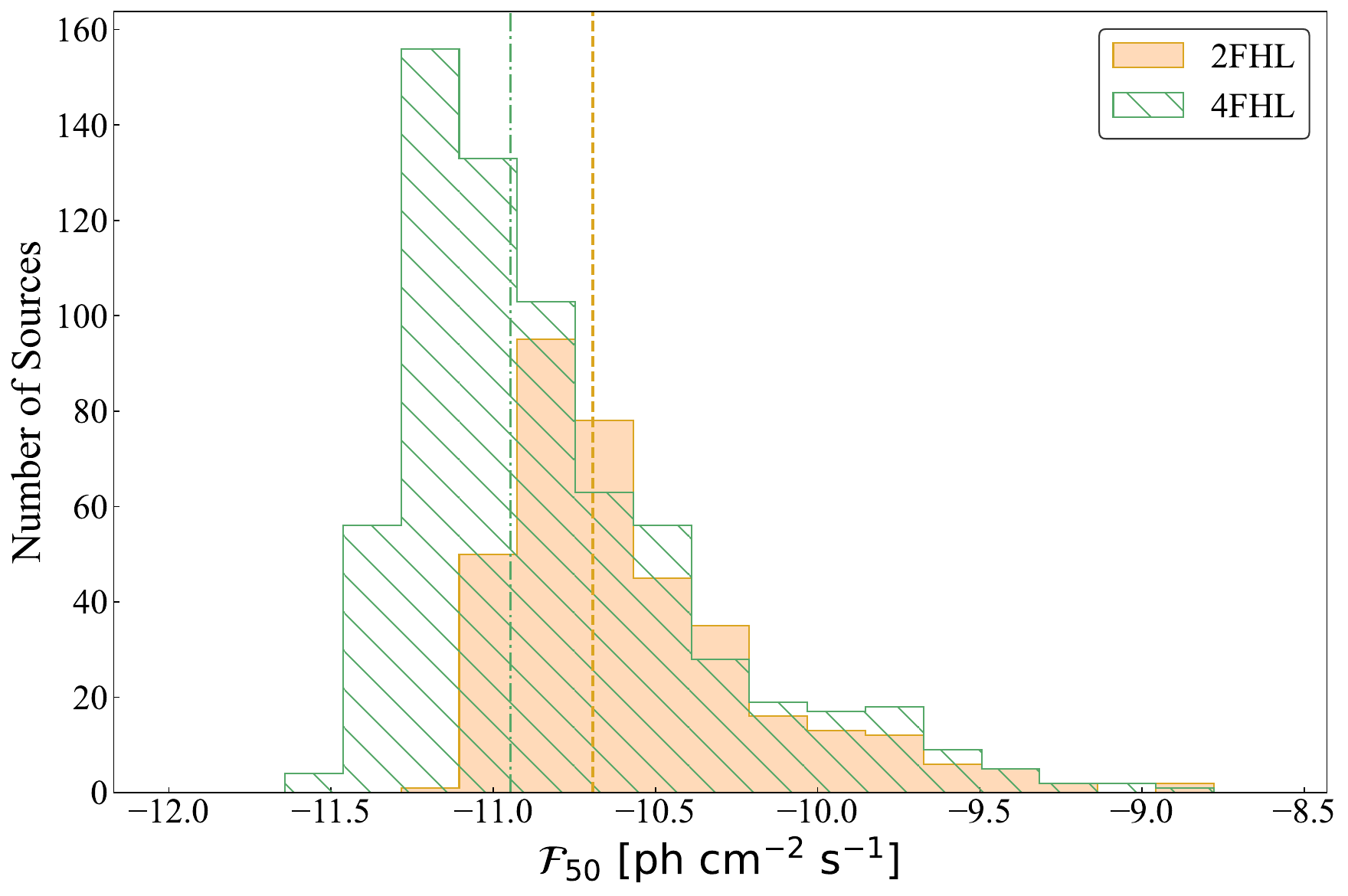}
    \caption{Flux distribution above 50 GeV comparing 2FHL (orange) and 4FHL (green slash) catalogs. The medians are marked with dashed and dash-dotted, respectively. The 4FHL catalog enhanced sensitivity allows for the detection of fainter sources.}
    \label{fig:flux_2fhl_4fhl}


    \includegraphics[width=0.6\textwidth]{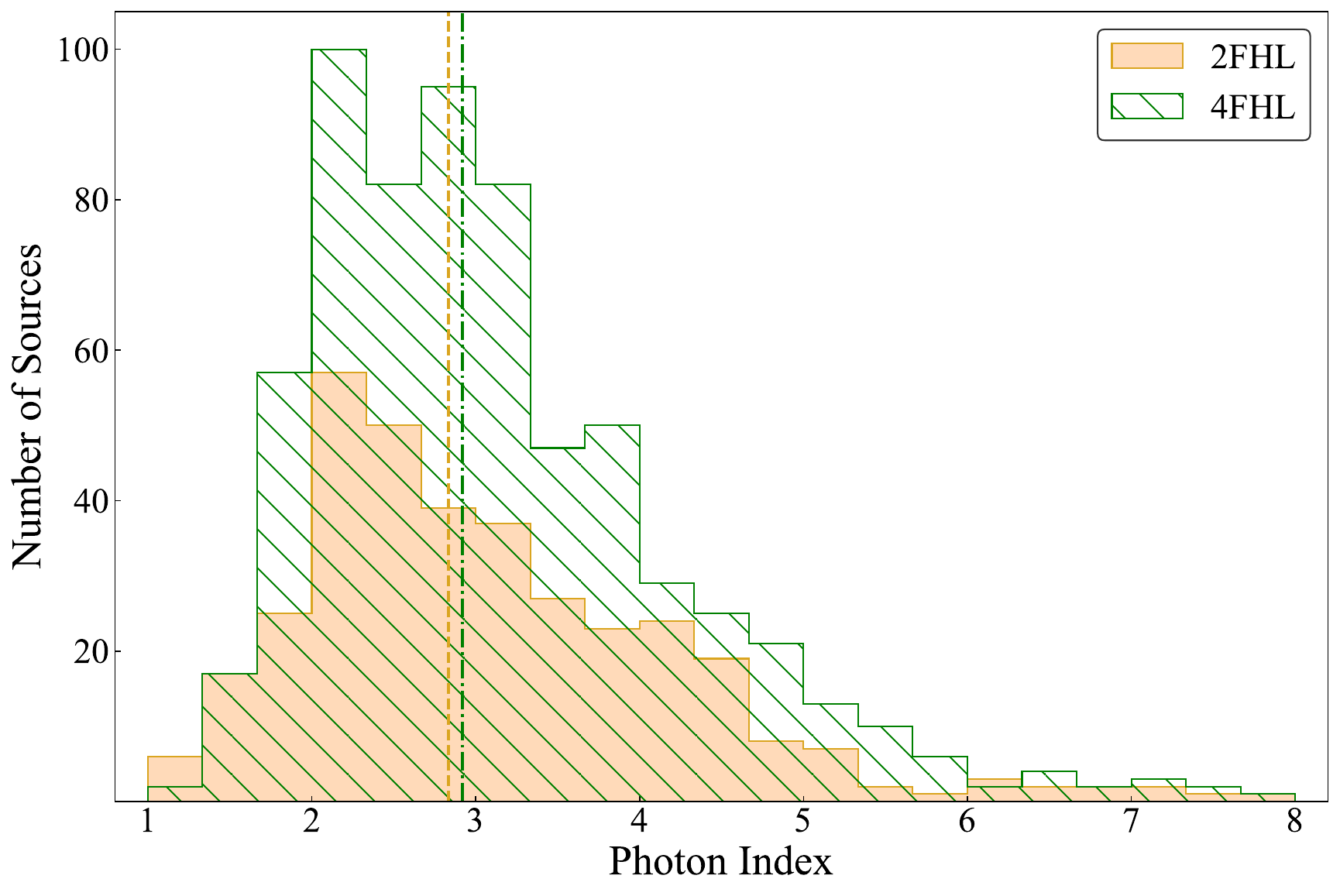}
    \caption{Observed photon index distribution above 50 GeV comparing 2FHL (orange) and 4FHL (green slash) catalogs. The medians are marked with dashed and dash-dotted, respectively. }
    \label{fig:spectral_distribution_comparison}

    \includegraphics[width=0.6\textwidth]{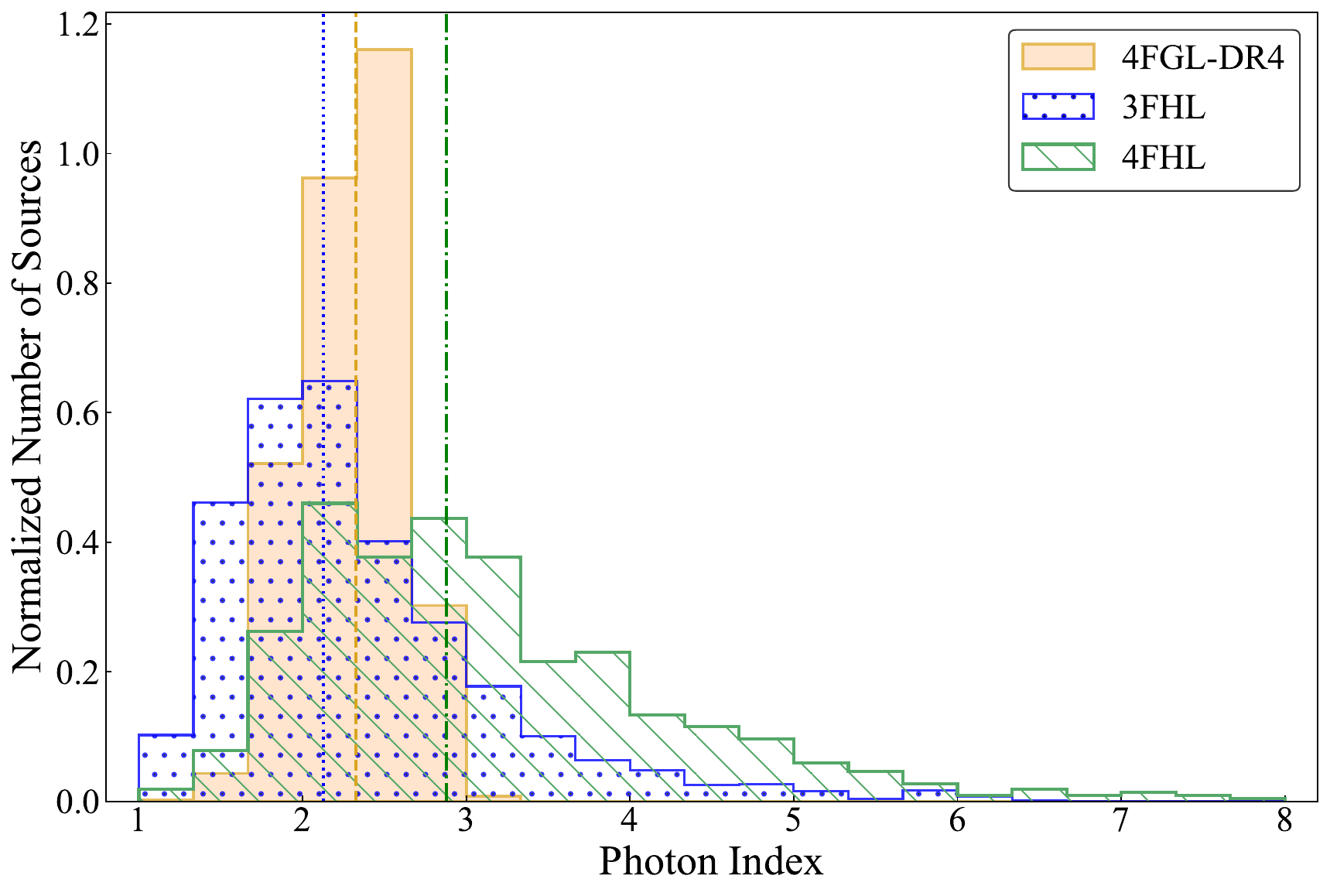}
    \caption{Normalized observed photon index distribution compared across 4FGL-DR4 (orange), 3FHL (blue dots), and 4FHL (green slash) catalogs. The medians are marked with dashed, dotted, and dash-dotted, respectively.}
    
    \label{fig:4fgl_3fhl_4fhl}
\end{figure*}

The sources in the 4FHL catalog have fluxes above 50\,GeV that range from approximately $2.9 \times 10^{-12}$ ph cm$^{-2}$ s$^{-1}$ to $1.2 \times 10^{-9}$ ph cm$^{-2}$ s$^{-1}$ (Crab Nebula), with a median flux of $1.1 \times 10^{-11}$ ph cm$^{-2}$ s$^{-1}$. This is a factor of 2 lower than 2FHL, which reports flux values between $8 \times 10^{-12}$ and $1.3 \times 10^{-9}$ ph cm$^{-2}$ s$^{-1}$ with a median flux of $2.0 \times 10^{-11}$ ph cm$^{-2}$ s$^{-1}$, enhancing the sensitivity to fainter sources. Figure \ref{fig:flux_2fhl_4fhl} shows the flux distribution of 2FHL and 4FHL.

The median observed photon index is $2.91$. The 1$\sigma$ uncertainty in the photon index increases notably for sources with larger photon indices. For instance, for sources with photon indices around $2$, the median uncertainty is approximately $0.45$. However, for sources with photon indices around $5$, the uncertainty increases significantly, with the median uncertainty reaching $1.68$. Only $8.5\%$ of the 4FHL sources have photon indices greater than $5$. Figure \ref{fig:spectral_distribution_comparison} shows that the observed photon index distribution is similar in the 2FHL and 4FHL catalogs, with an increased number of softer sources around $\Gamma \sim 3$ (mostly extragalactic). Figure \ref{fig:4fgl_3fhl_4fhl} shows that the 4FHL median index, measured in the 50 GeV-2 TeV band, is shifted towards larger indices compared to those in 3FHL and 4FGL-DR4, reflecting the 4FHL dominance of extragalactic sources for which the observed spectrum above 50 GeV generally samples the declining part of the spectral energy distribution (SED). Note that, because 4FHL is built from sources detected above 50 GeV, the 4FGL-DR4 counterparts of 4FHL sources correspond to the harder subset of the 4FGL-DR4 population, with a median power-law index of 1.91 compared to 2.32 for the full 4FGL-DR4 catalog. The scatter of the index distribution also increases at higher energies, which is a statistical effect arising from the reduced number of detected photons above 50 GeV relative to 4FGL-DR4 and 3FHL.

Regarding the position uncertainty, half of the sources in the 4FHL catalog are localized to better than $1.47'$ in radius at 68\% confidence, corresponding to an improvement of about 14\% compared to the 2FHL median ($1.7'$). For sources with counterparts in 2FHL, the source localization has improved by about a factor of 2. Note that the systematic correction factors used were larger for 4FHL than 2FHL.

Thirty-five sources reported in 2FHL are missing in 4FHL. All of these sources were part of the seed list in the 4FHL analysis pipeline but were not reported in the final catalog since their 4FHL TS values fall below the detection threshold (TS$=25$), and their 2FHL TS values were also generally close to that threshold. Only one of them, 2FHL J1834.6$-$0701, had been significantly detected in 2FHL with TS$= 46$. That source is inside the SNR G024.7+00.6, which was included as an extended source in our pipeline. The fact that these other sources were not significant in 4FHL may be attributed to the use of improved background models or to intrinsic source variability.

\begin{figure}[htbp]
\centering
\includegraphics[width=1\textwidth]{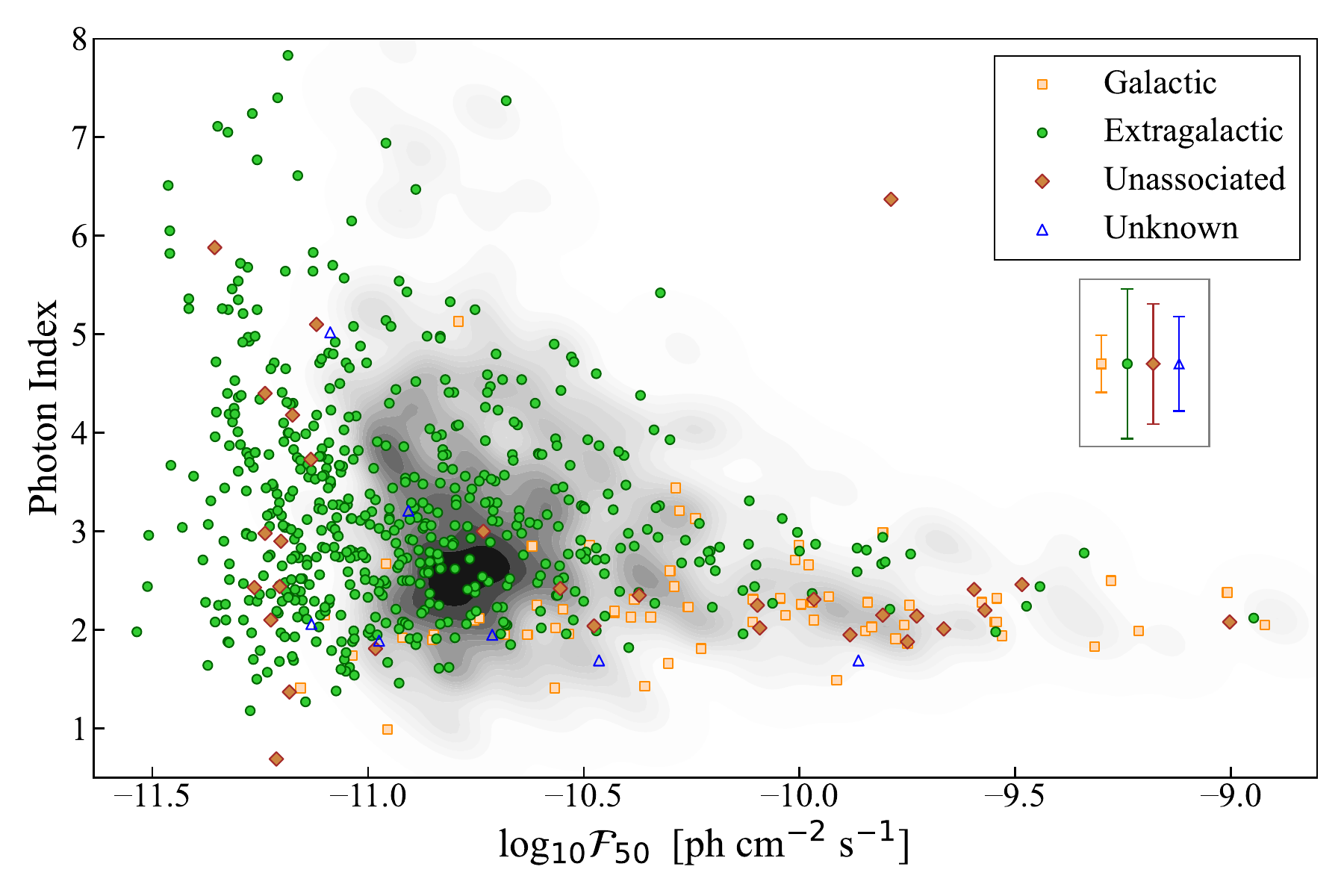} 
\caption{Observed photon index as a function of integrated photon flux for sources detected above 50 GeV in the 4FHL catalog. Galactic sources (orange squares), extragalactic sources (green circles), and unassociated sources (brown diamonds) are shown. The density contour in grayscale represents the 2FHL distribution for comparison. The median uncertainty in the photon index is shown for each case as error bars, with Galactic sources having a smaller uncertainty. The median flux uncertainty is $1.1 \times 10^{-11}$  ph cm$^{-2}$ s$^{-1}$ for Galactic sources, $3.9 \times 10^{-12}$  ph cm$^{-2}$ s$^{-1}$ for extragalactic sources and $6.3 \times 10^{-12}$  ph cm$^{-2}$ s$^{-1}$ for unknown/unassociated sources. The larger median flux uncertainty for Galactic sources reflects the higher diffuse background in the Galactic plane.}
\label{fig:index_flux}
\end{figure}

Figure \ref{fig:index_flux} shows the photon index versus the integrated photon flux above 50 GeV for Galactic objects,  sources associated with extragalactic nature, unassociated sources, and sources of unknown $\gamma$-ray population.  The detection threshold for photon flux does not show a strong dependence on the photon index, mainly because above 50 GeV the analysis is close to photon-limited, with a low diffuse background contribution and a nearly constant PSF above 50 GeV. Extragalactic sources are detected at lower flux levels than Galactic sources, as expected given the anisotropic diffuse background that reduces sensitivity in the Galactic plane. 

\begin{figure*}[htbp]
    \centering
    
    \includegraphics[width=0.8\textwidth]{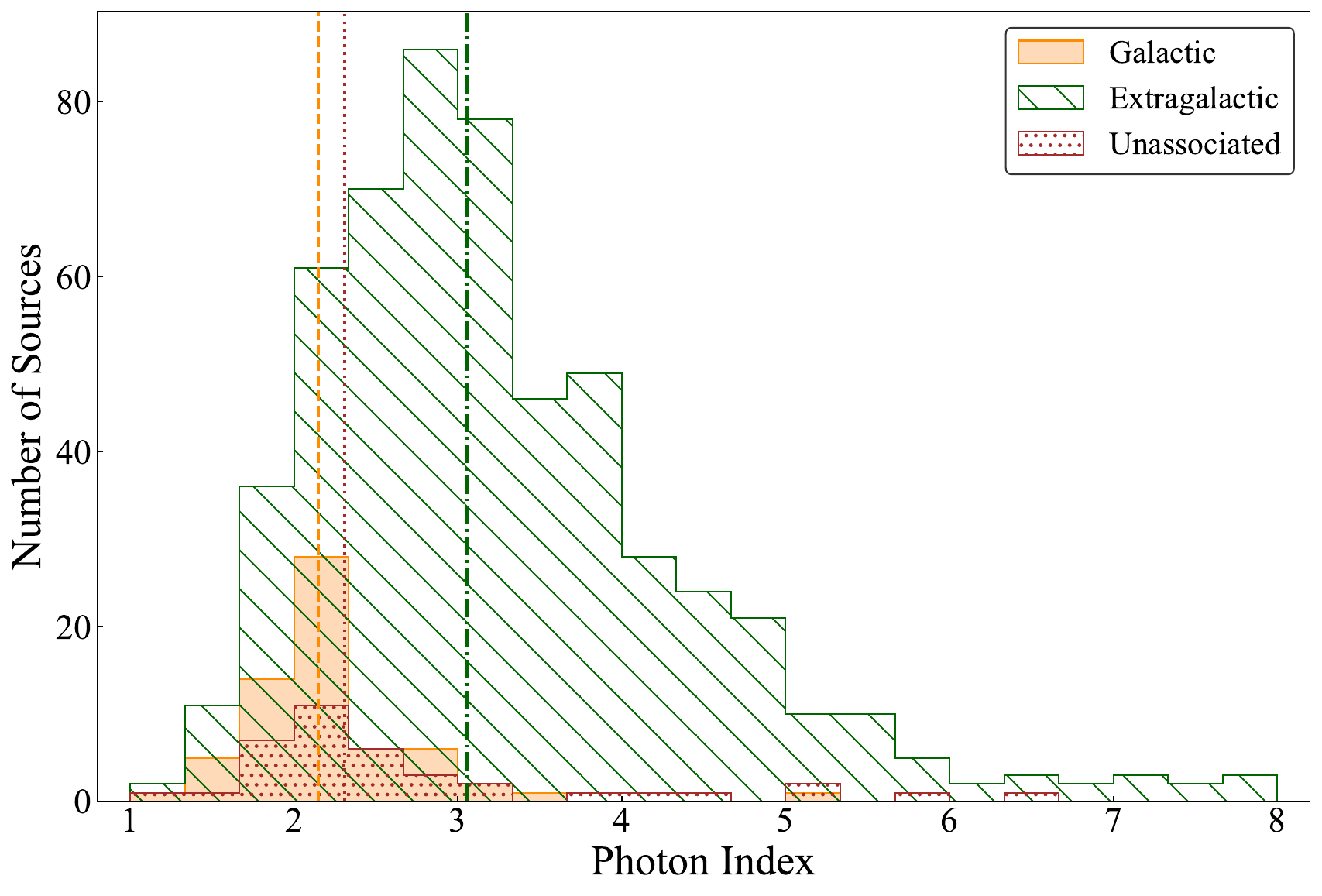}
        \caption{Observed photon index distribution for Galactic sources (orange), extragalactic sources (green), and unassociated sources (brown). The Galactic population exhibits harder observed spectra, with the median value being significantly lower than that of the extragalactic sources. The medians are marked with dashed, dash-dotted, and dotted vertical lines, respectively.}
        \label{fig:spec_dist}

    \vspace{0.8cm}
    \includegraphics[width=0.8\textwidth]{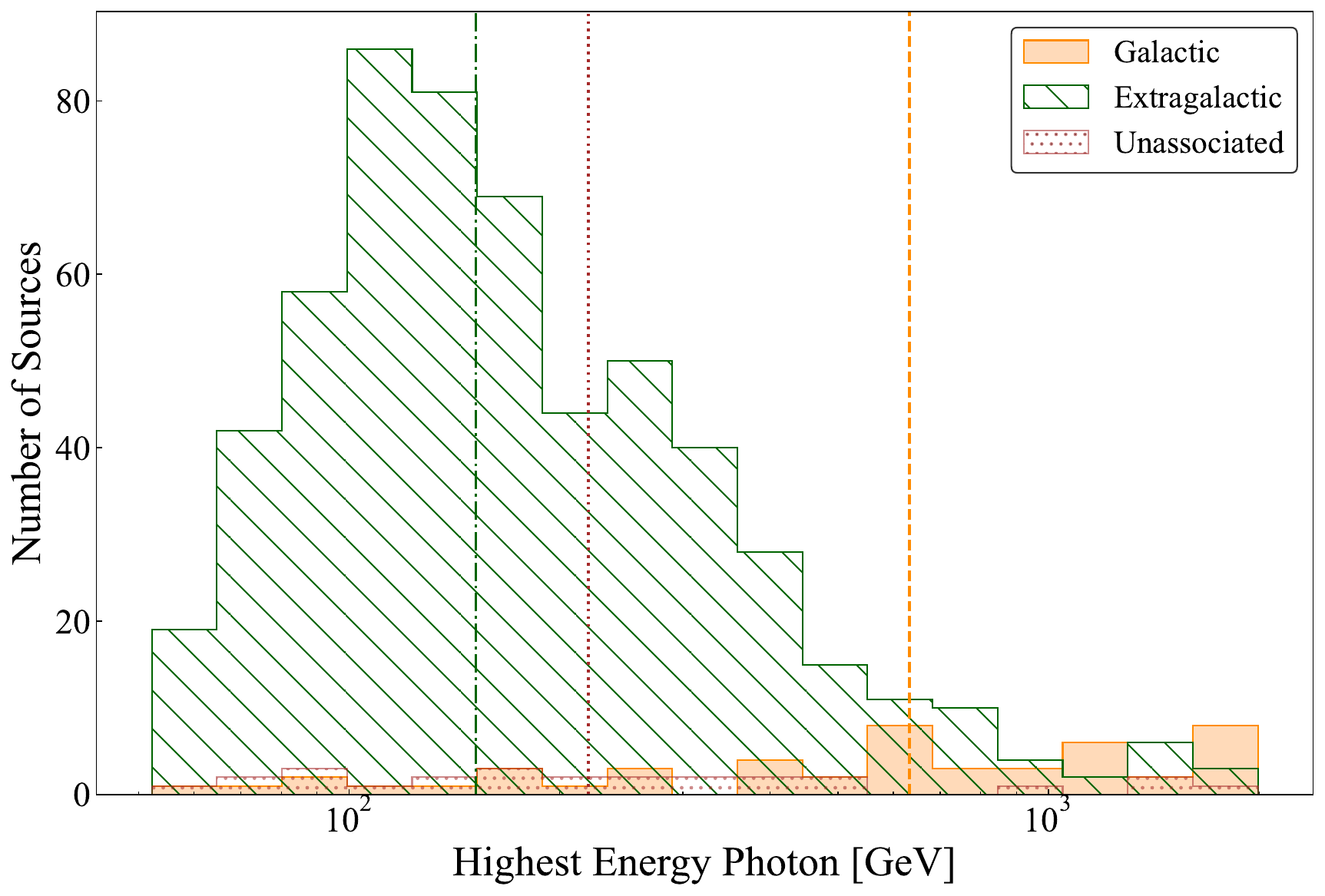}
        \caption{High-Energy Photon (HEP) distribution for Galactic (orange), extragalactic (green), and unassociated (brown) sources.
        Galactic sources have a higher median HEP value, consistent with their harder observed spectra, while for extragalactic sources the gamma-ray horizon limits the maximum detectable photon energy.
        }
        \label{fig:hep_dist}
    \label{fig:}
\end{figure*}

Figures \ref{fig:spec_dist} and \ref{fig:hep_dist} show the distributions of the observed photon index and the HEP according to different source populations. Galactic sources typically exhibit harder observed spectra, whereas extragalactic sources tend to have softer observed spectra spanning a broader range of indices, peaking around $\Gamma \sim 2.5$ at $>50$ GeV. For extragalactic sources, the soft observed indices may partly reflect attenuation by the extragalactic background light (EBL), depending on redshift. 
Galactic sources generally reach higher observed HEP values.  In particular, 31 sources reach HEP $>1$ TeV, of which 17 are associated with Galactic origin, 11 with extragalactic sources, while three remain unassociated (4FHL J0810.9+3533, 4FHL J1729.9$-$3336 and 4FHL J1745.8$-$3028e).

\subsection{Galactic Source Population} \label{sec:Galactic}
\begin{figure*}
\vspace{-0.9cm}
    \centering
    \begin{subfigure}{1\textwidth}
        \includegraphics[width=1.\linewidth]{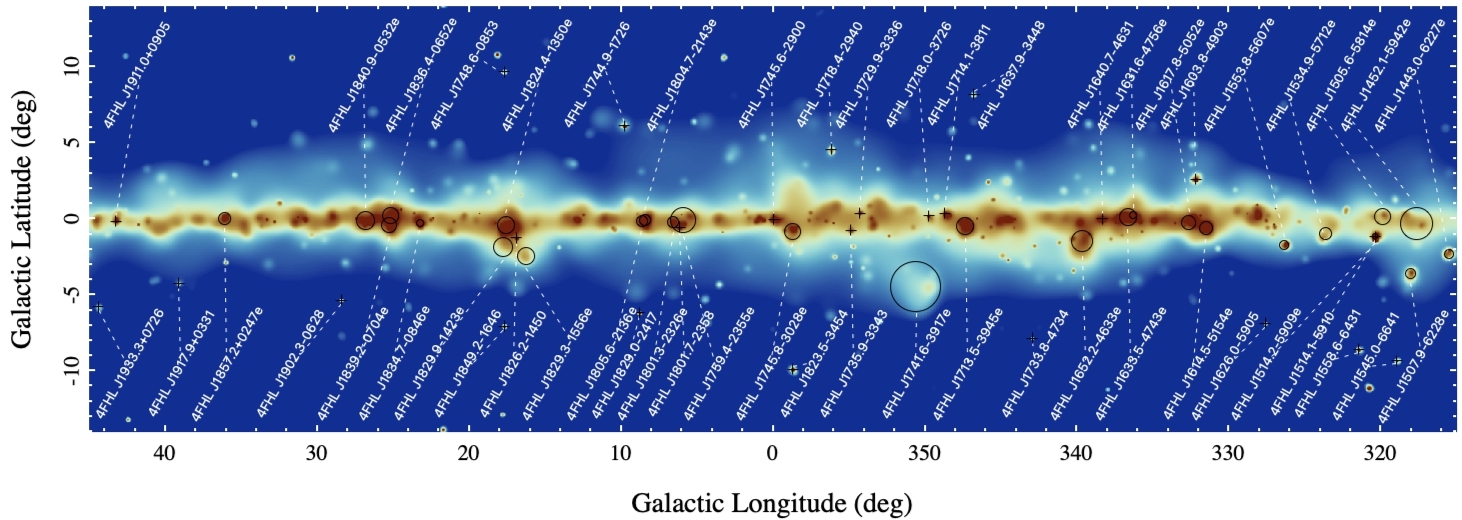}
    \end{subfigure}
    \vspace{-0.95cm} 
    
    \begin{subfigure}{1\textwidth}
        \includegraphics[width=\linewidth]{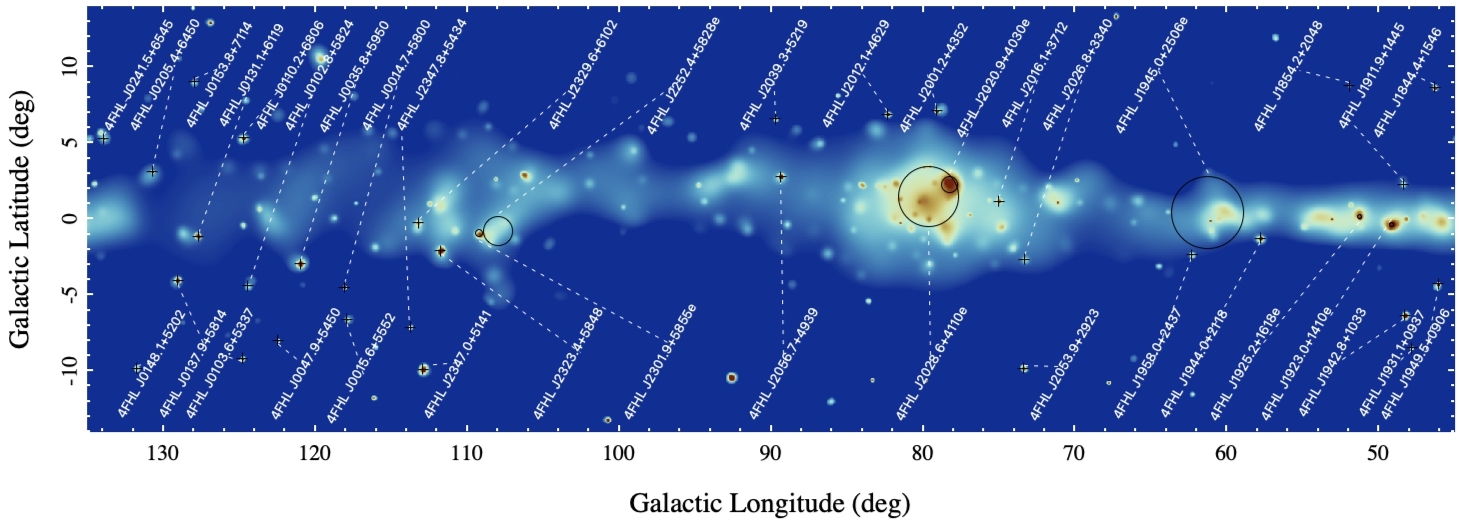}
    \end{subfigure}
    \vspace{-0.95cm}
    
    \begin{subfigure}{1\textwidth}
        \includegraphics[width=\linewidth]{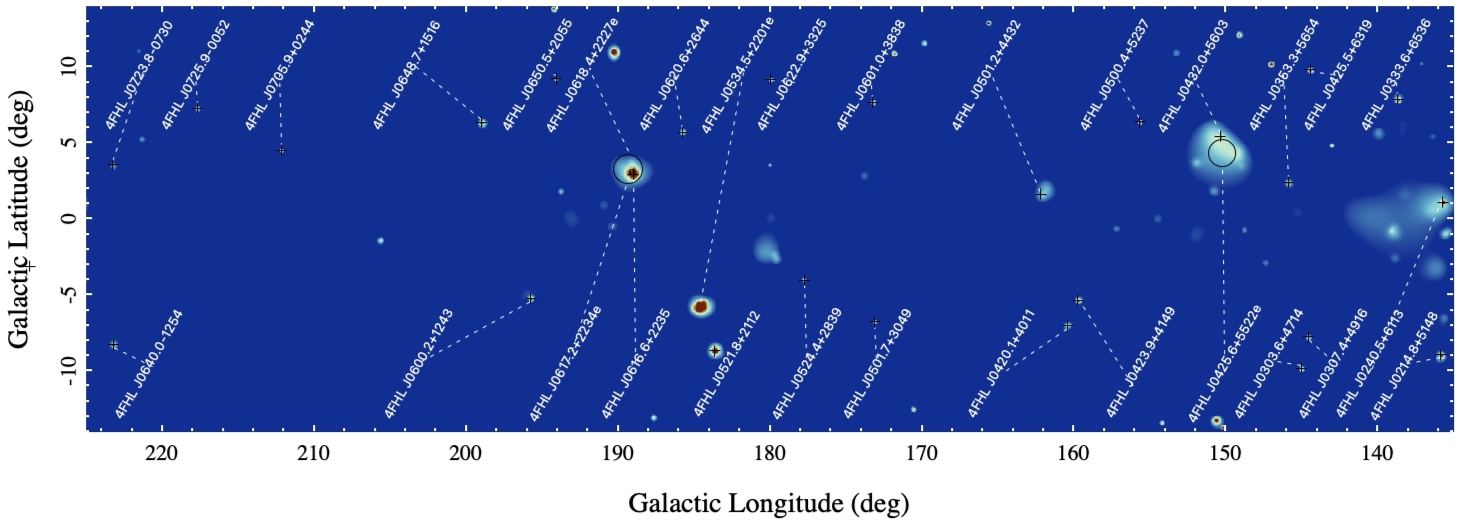}
    \end{subfigure}
    \vspace{-0.95cm}
    
    \begin{subfigure}{1\textwidth}
        \includegraphics[width=\linewidth]{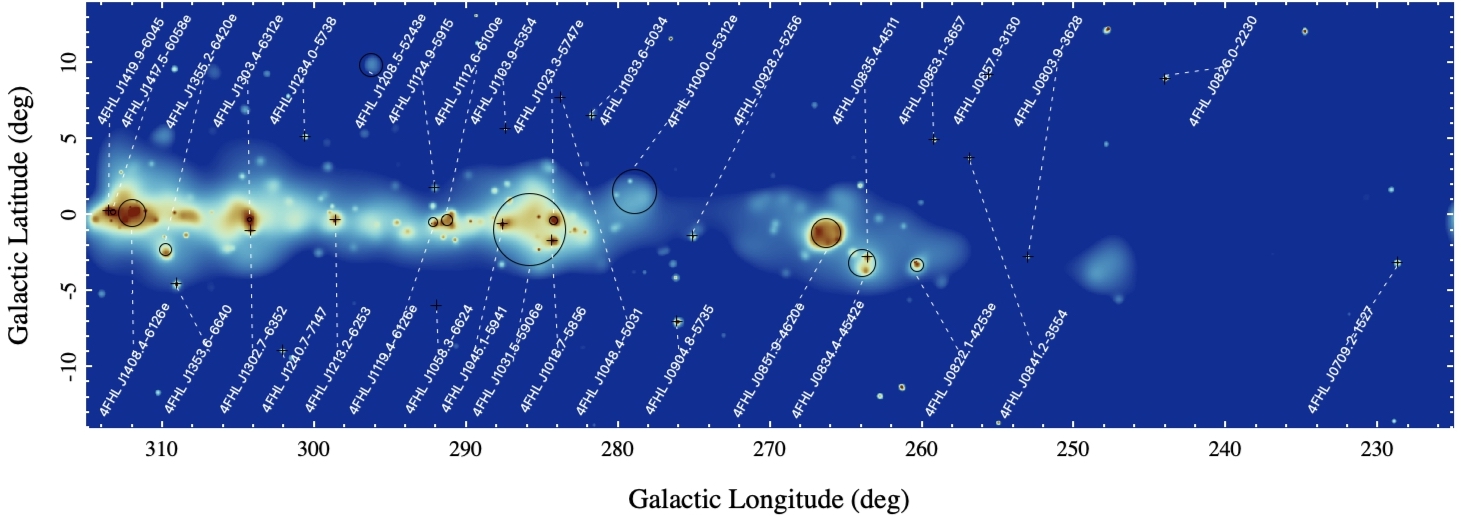}
    \end{subfigure}
    
    \caption{Adaptively smoothed count map of the Galactic plane, shown in four panels for the latitude range $-14^{\circ} \leq b \leq +14^{\circ}$. The panels are centered at $l=0^{\circ}$, $90^{\circ}$, $180^{\circ}$, and $270^{\circ}$. Only sources detected at $-10^{\circ} \leq b \leq +10^{\circ}$ are labeled. Point-like sources are marked with crosses, while extended sources are indicated with circles whose sizes reflect their extensions.}
    \label{fig:fourpanels}
\end{figure*}

The narrow PSF core of the LAT, with an angular resolution of approximately 0.1° above 50 GeV, allows for the effective study of sources in the plane of our Galaxy. In the 4FHL catalog, within $|b|< 10^{\circ}$, a total of 156 sources were detected. This is an increase relative to 2FHL, which reported 103 sources in the same region. Among the 156 low-latitude detections, 60 are associated with Galactic sources, 73 with blazars, one radio galaxy (TXS 0149+710), and 22 remain unassociated or unknown. Figure \ref{fig:fourpanels} shows cut-outs of the Galactic plane, where all sources detected within $|b| < 10^{\circ}$ are labeled. Note that adaptive smoothing can make sub-threshold photon clusters appear source-like. Several of these structures were already present in the 2FHL Galactic maps. A clear example is the point-like source 4FHL J0904.8$-$5735, which appeared as a fluctuation in 2FHL and is now detected in 4FHL.

Regarding the 60 (+5 located at $b\geq 10^{\circ}$) Galactic sources, a notable fraction of them are associated with high-energy phenomena such as SNRs, PWNe, and X-ray binaries. 
In particular, we identify 21 SNRs, 16 PWNe, one pulsar (PSR; Vela Pulsar), four high-mass X-ray binaries (HMBs), one binary system (BIN), the galactic center, and two star-forming regions (SFRs), corresponding to Westerlund 2 and the Cygnus X cocoon, as well as 18 sources classified as SNR/PWN candidates (SPP). 
These results are consistent with the detections in the 2FHL catalog, where SNRs and PWNe also dominated the Galactic source population, suggesting that the Galactic sources detected above 50 GeV are associated with objects in the late stages of stellar evolution. 

\begin{figure}[htbp]
    \centering
    \begin{minipage}{0.45\textwidth}
        \centering
        \includegraphics[width=\linewidth]{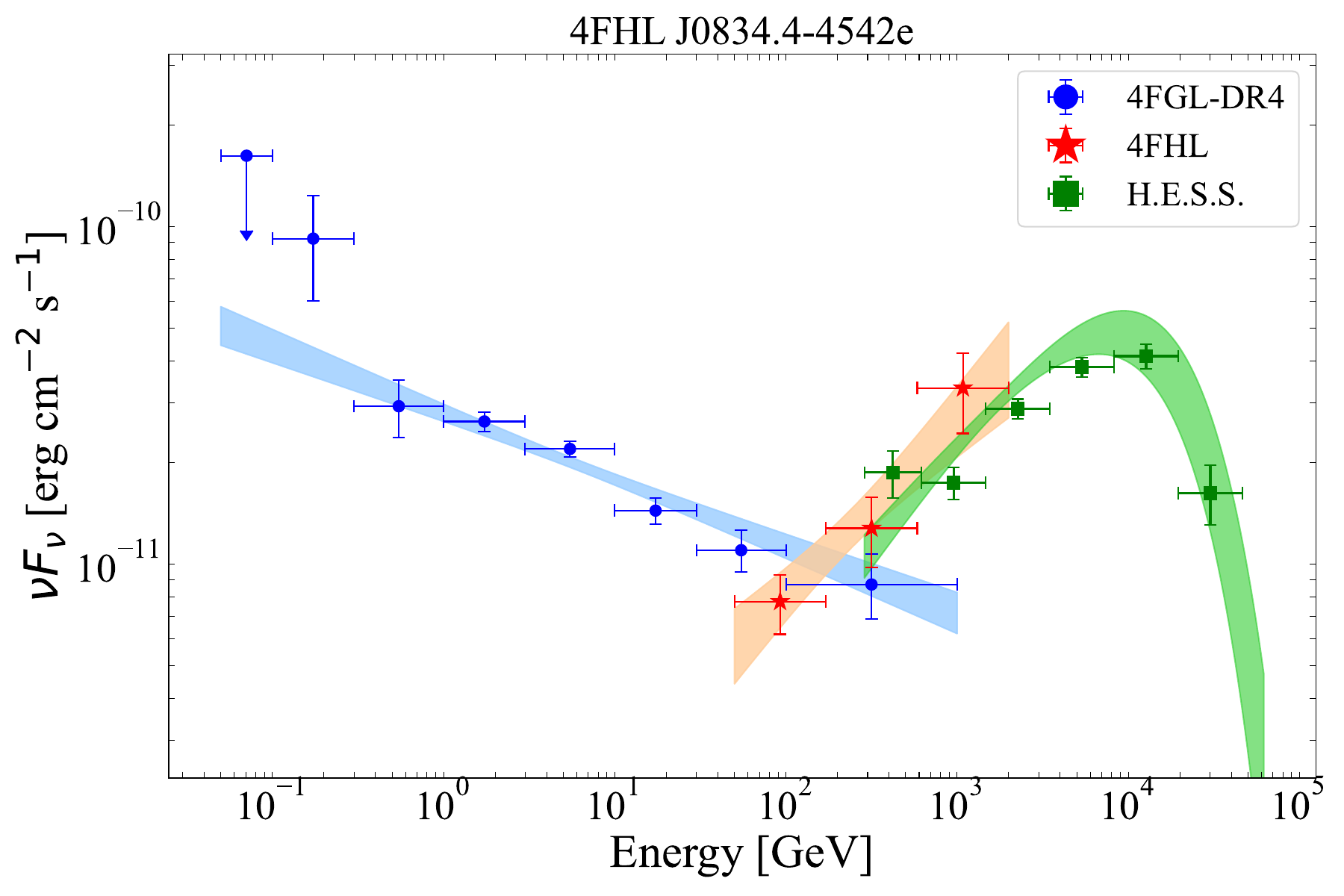} 
    \end{minipage}
    \hspace{0.05\textwidth} 
    \begin{minipage}{0.45\textwidth}
        \centering
        \includegraphics[width=\linewidth]{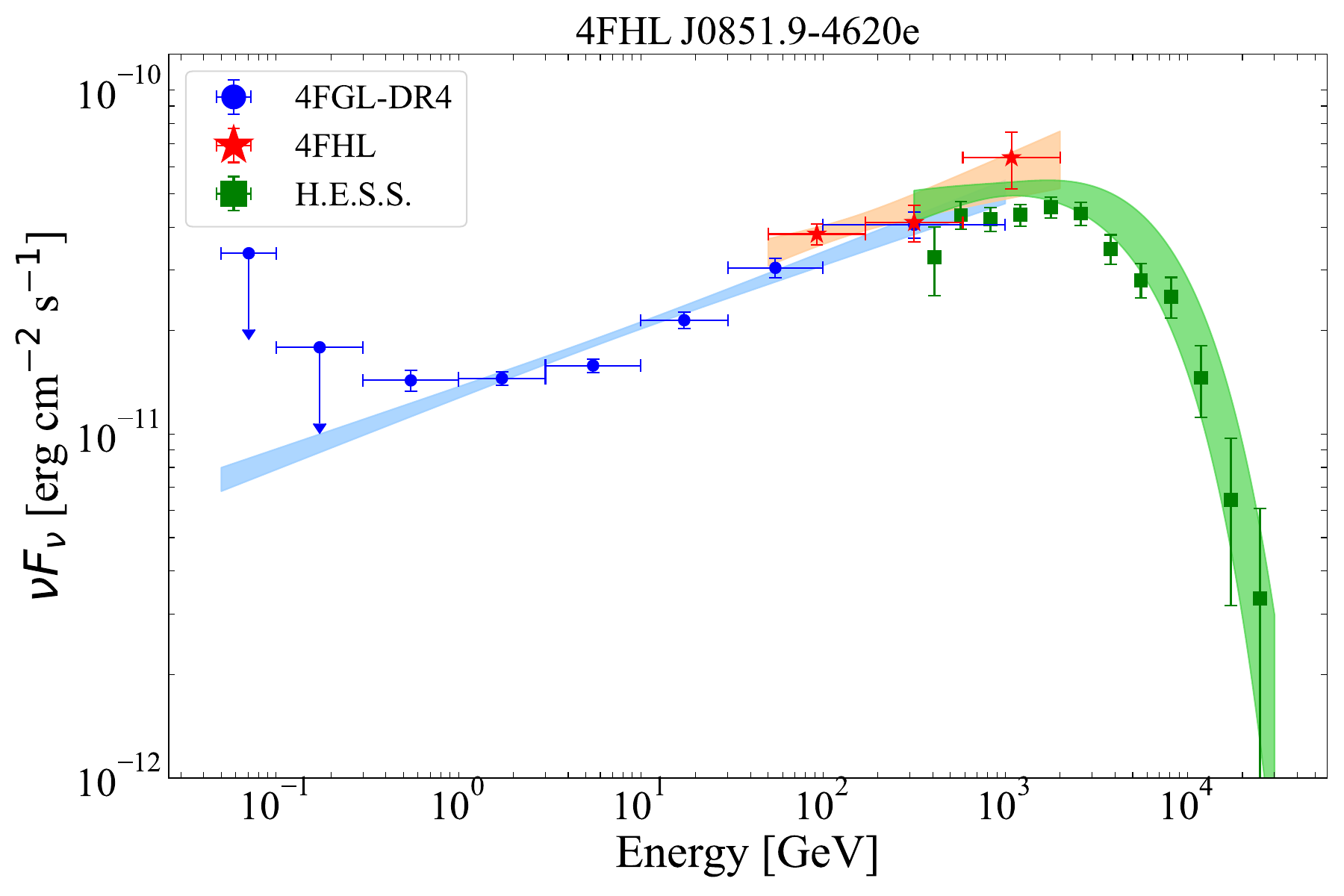}
    \end{minipage}

    \begin{minipage}{0.45\textwidth}
        \centering
        \includegraphics[width=\linewidth]{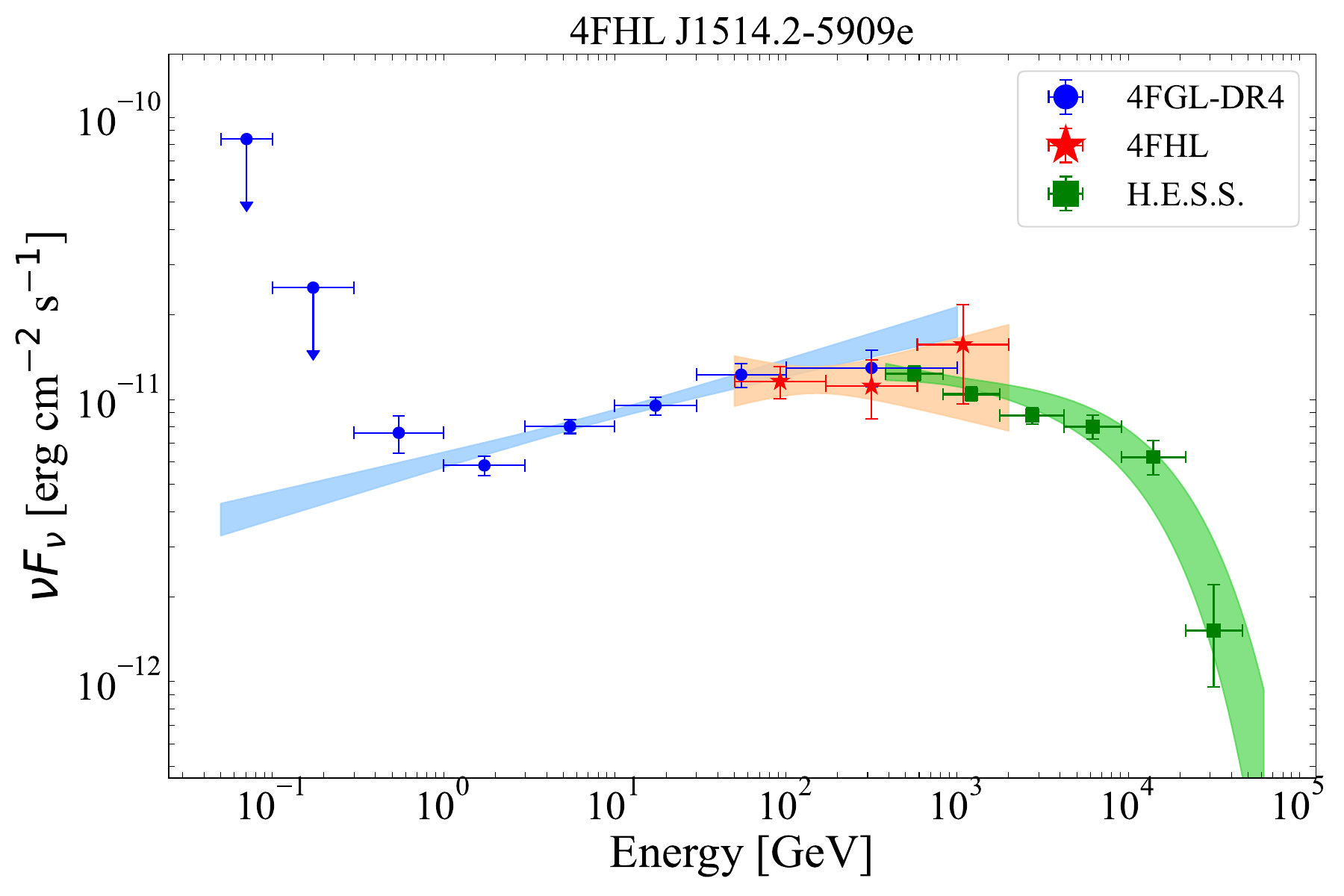} 
    \end{minipage}
    \hspace{0.05\textwidth} 
    \begin{minipage}{0.45\textwidth}
        \centering
        \includegraphics[width=\linewidth]{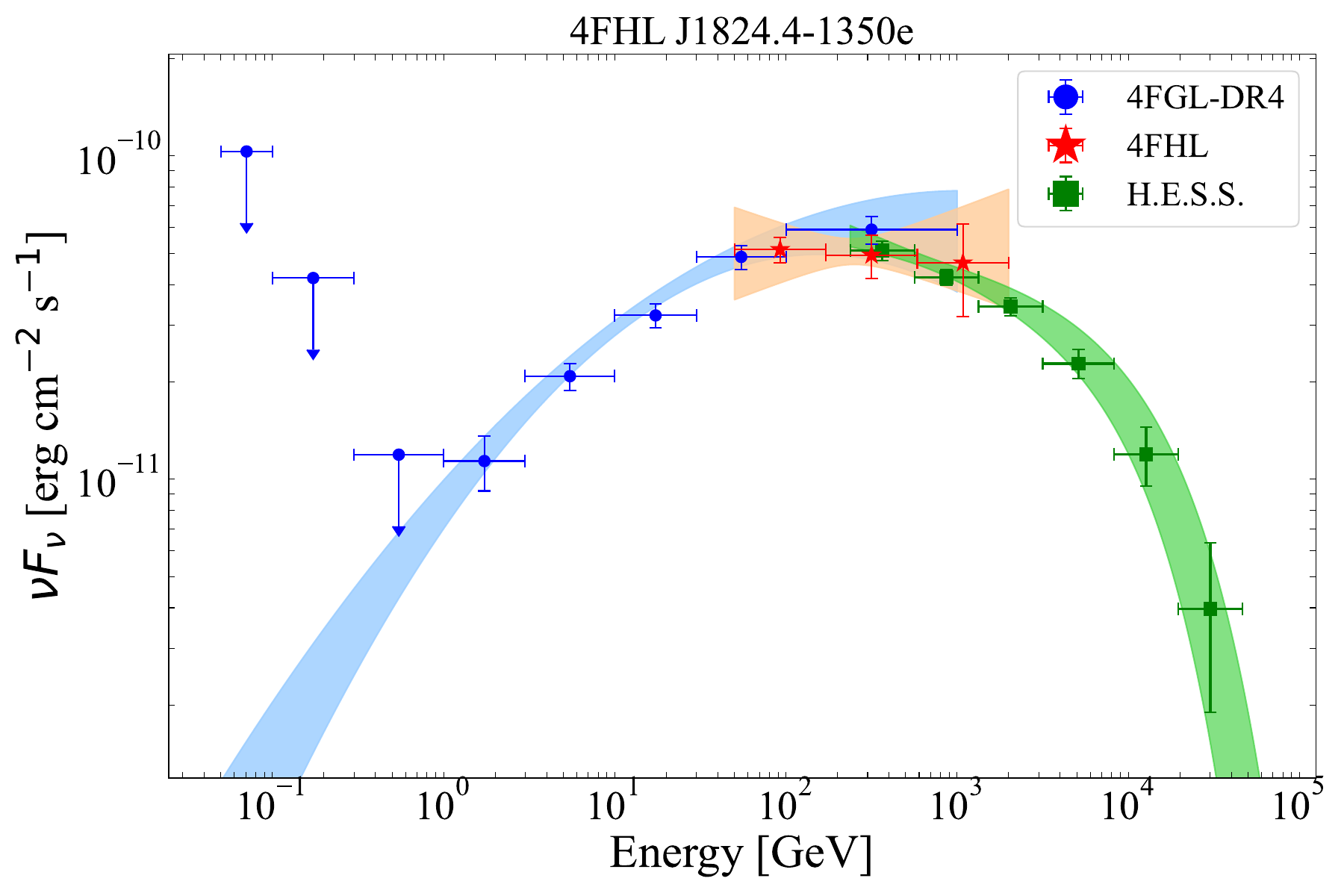} 
    \end{minipage}

    \begin{minipage}{0.45\textwidth}
        \centering
        \includegraphics[width=\linewidth]{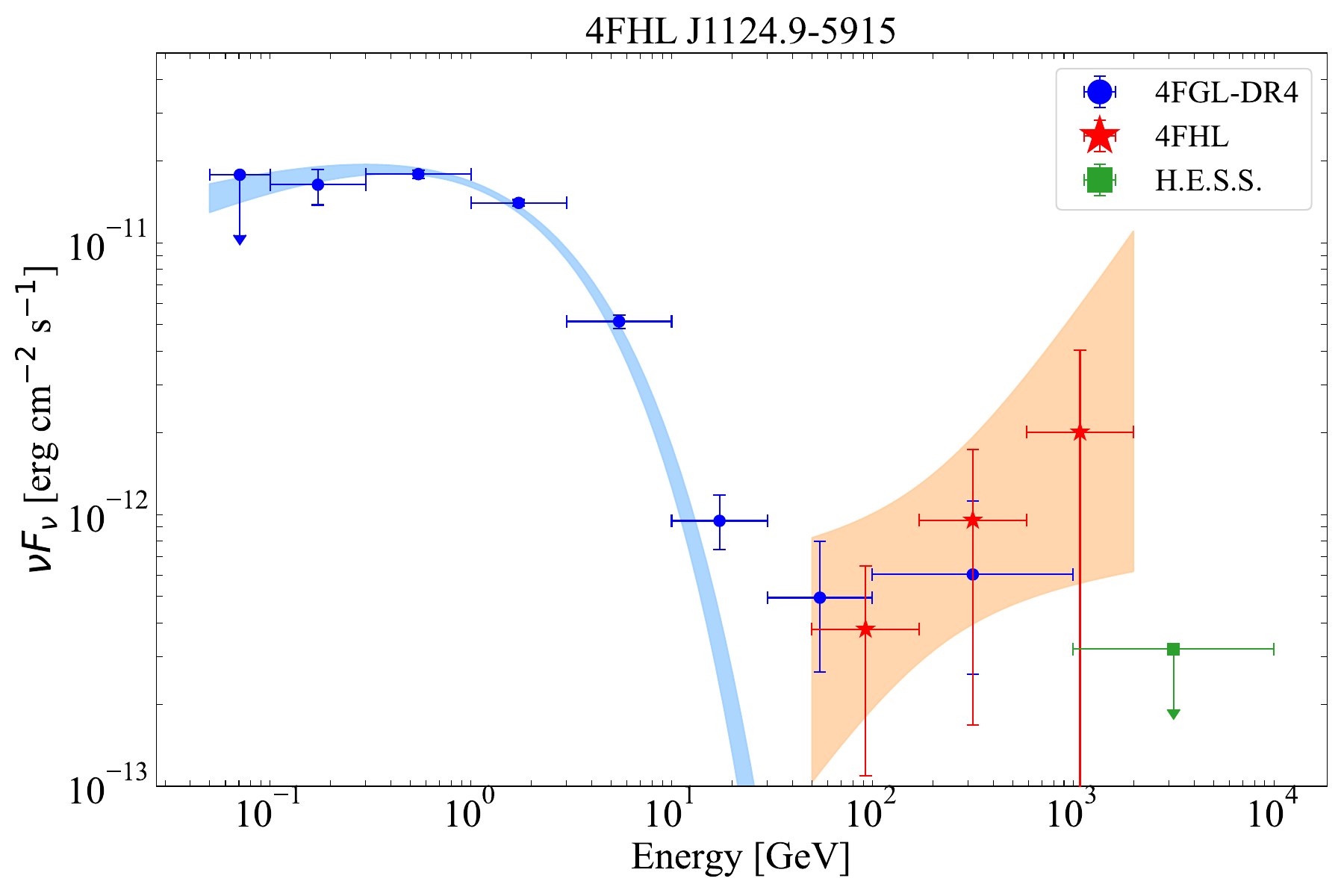} 
    \end{minipage}

    \caption{Spectral energy distributions of five Galactic sources discussed in \S \ref{HESS}: (top left) PWN Vela X, (top right) SNR Vela Junior, (medium left) PWN MSH 15-52, and the PWN powered by PSR J1826$-$1334 (medium right). 
    The bottom panel presents  sources not detected by H.E.S.S., with 95\% upper limits adopted from \citet{HESSUL}. 
    The data include 4FGL-DR4 (blue dots), 4FHL (red stars),  and H.E.S.S. (green squares).}
    \label{fig:sed_sources}
\end{figure}

Galactic sources typically show hard observed spectra, with a median photon index of 2.15 (see Figure~\ref{fig:spec_dist}). Such hard spectra are consistent with SEDs peaking at $\sim$ 1 TeV or beyond, as we can see for the PWN Vela X in Figure \ref{fig:sed_sources}. This is a clear sign of the presence of energetic particles accelerated by efficient mechanisms \citep{abdo2011}. 
PWNe are harder than SNRs above 50 GeV, with 40\% of 4FHL PWNe having an observed photon index $\leq 2.0$. All PWNe detected by {\it Fermi}-LAT to date are powered by young and energetic pulsars \citep[$\tau \lesssim 30$ kyr;][]{acero2013}. An example is shown in Figure~\ref{fig:sed_sources}, which presents the SED of the PWN MSH 15-52 (4FHL J1514.2$-$5909e), powered by the young \citep[age 1.7 kyr;][]{PSRB1509-58} high-magnetic field pulsar PSR B1509$-$58.
Regarding SNRs, 33\% of those detected in the 4FHL have a photon index $\leq 2.0$. Such hard spectra are typically characteristic of young or middle-aged remnants in tenuous environments, which can be difficult to identify, even in radio surveys. Consequently, some of the new or unidentified 4FHL sources may correspond to previously unrecognized young SNRs, such as 4FHL J0501.2+4432 or 4FHL J1729.9$-$3336, further discussed in \S~\ref{sec:new_gal}.

Of the 22 sources that remain unassociated/unknown, 11 have power-law indices of $\Gamma \leq 2.0$, suggesting a likely Galactic origin, since only $\sim8\%$ of 4FHL blazars above 50~GeV have $\Gamma \le 2$, as their emission is typically observed beyond the IC peak and softened by the EBL absorption \citep[e.g.][]{Dominguez_2011,EBL_saldana_2021,EBL_Finke_2022}.
Fifteen were modeled as extended, with eleven of these included in the recently released 2FGES catalog. A more detailed discussion of the extended sources is provided below. 

Among the point-like unassociated detections, 4FHL J1854.2+2048 is a hard ($\Gamma= 1.37\pm0.48$) and faint ($F_{50}$ = $6 \times 10^{-12}$ ph cm$^{-2}$ s$^{-1}$) source that remains unassociated, including in 4FGL-DR4. This unusual combination of hardness and faintness makes it a promising target for further multi-wavelength observations.

\subsubsection{Extended Sources}\label{sec:extended}
In total, 103 sources were modeled as spatially extended and included in the ML analysis. Unlike in 2FHL, no dedicated search for new spatially extended sources or test for possible extension of sources detected as point-like were performed in this work. We included the 35 non-dubious extended sources from 2FGES \citep{2FGES} and the 4FGL-DR4 extended sources that do not overlap (by more than $0.2^\circ$) with 2FGES objects.

After performing the analysis, 55 extended sources were detected in 4FHL, representing more than half of the known extended  {\it Fermi} sources entered in the analysis. This marks a notable increase compared to the 2FHL catalog, which reported 36 extended sources, all of which are also detected in 4FHL.

Out of the 35 extended sources included in the 2FGES catalog, 23 were detected with 11 of them remaining unassociated or of unknown type, corresponding to nearly 50\% of the Galactic population that remain unidentified. According to the possible associations listed in the 2FGES catalog, 
4FHL J1617.8$-$5052e was identified as PWN, while 4FHL J1452.1$-$5942e and 4FHL J1829.3$-$1556e were classified as SPP sources. 4FHL J1505.6$-$5814e is located near the unidentified TeV source HESS J1503$-$582 (angular separation of $0.3^{\circ}$) and the blazar 4FGL J1503.7$-$5801. Recent studies \citep{HESSSNR} suggest that the blazar emission may contribute to the TeV emission of HESS J1503$-$582. Nevertheless, its origin remains uncertain. Possible interpretations include an old SNR with a slowly expanding shell, a pulsar halo, or the combined activity of stellar winds and supernova explosions from massive stars in a nearby OB association \citep{OBHESS,HESSSNR}. The remaining sources (4FHL J1034.0$-$5832e, J1112.6$-$6100e, J1408.4$-$6126e, 4FHL J1417.5$-$6058e, J1759.4$-$2355e, J1829.9$-$1423e, 4FHL J1945.0+2506e) were reported as unassociated or unknown. In particular, 4FHL J1112.6$-$6100e was also detected in 2FHL and associated with the pulsar PSR J1112$-$6103, which partially overlaps with the massive star forming region NGC 3603.

Of the 12 2FGES sources not included in 4FHL, three had $10<\mathrm{TS}<25$ and were therefore not reported in the final catalog. 


Regarding the 68 sources modeled as extended in 4FGL-DR4, 32 are significantly detected above 50 GeV, while seven remain unassociated. For the Large Magellanic Cloud (LMC), we included the four independent contributions (LMC-FarWest, LMC-Galaxy, LMC-30DorWest, and LMC-North); however, only the harder components LMC-30DorWest and LMC-North are significantly detected above 50 GeV. The Crab Nebula is modeled as extended ($R_{68}=0.03^\circ$, \citeauthor{Ackermann_2018_extended}~\citeyear{Ackermann_2018_extended}) using the inverse Compton (IC) component, which is detected in 4FHL.  

In total, 36 extended 4FGL-DR4 sources remain undetected in 4FHL. 
In the Cygnus region, the Cygnus Cocoon is detected, whereas the extended SNR Cygnus Loop is not, as expected given its soft spectrum above 10 GeV \citep{Tutone2021}. A relatively bright BL Lac, 4FHL J2053.9+2923, is detected close to the south-western border of the remnant.

Six extended sources reported in 4FGL-DR4 are not detected as extended in 4FHL, but instead appear as point-like detections or have nearby ($<0.2^{\circ}$ separation) point-like counterparts. In particular, the PWN 3C~58 is not detected as extended emission above 50 GeV, instead, a point-like source (4FHL J0205.4+6450) is detected. Regarding the lobes of Centaurus~A, we detect the radio galaxy (4FHL J1325.5$-$4259) together with an associated blazar (4FHL J1315.1$-$4236) inside the lobes. The SNR S147 is not detected either in the 2FHL or 3FHL catalogs.


Finally, 4FHL J0425.6$+$5522e is associated with the SNR G150.3$+$04.5 and is spatially coincident with 1LHAASO J0428$+$5331, which has been proposed as a potential cosmic-ray PeVatron candidate. \cite{li2024evidencehybridgammarayemission} indicate a hybrid emission scenario for this system, with leptonic processes favored in the northern lobe and hadronic interactions in the southern lobe. However, \citet{Devin2020} suggest that the southern source is more consistent with a pulsar origin. In addition, we report a new LAT hard point-like source, 4FHL J0432.0$+$5603 ($\Gamma=1.74 \pm 0.44$), located in the northern lobe of the SNR.

\subsubsection{Comparison with the H.E.S.S. Galactic Plane
Survey
}\label{HESS}

The High Energy Stereoscopic System (H.E.S.S.) has conducted an extensive survey of part of the Galactic plane at TeV energies. With a field of view (FOV) of about $5^{\circ}$ and an angular resolution of $\sim0.12^{\circ}$, the H.E.S.S. array accumulated nearly 2800 hours of exposure, reaching an average sensitivity of $\sim$1.5\% of the Crab Nebula flux ($2.3 \times 10^{-11}$ ph cm$^{-2}$ s$^{-1}$) at energies above 1 TeV \citep{Aharonian_2006}, detecting 78 sources as reported in the latest Galactic sky catalog \citep[HGPS;][]{HESS}. The survey primarily covers the inner Galaxy, spanning Galactic longitudes from $250^{\circ}$ to $65^{\circ}$ and latitudes within $|b| < 3.5^{\circ}$. For comparison, we estimate that the average LAT sensitivity above 50 GeV in the same region covered by HGPS corresponds to $1\%$ of the 50\,GeV$–$2\,TeV Crab Nebula flux ($10^{-11}\ \mathrm{ph\ cm^{-2}\ s^{-1}}$).

While both surveys overlap, their sensitivity to different energy ranges provides a complementary view of the very high-energy (VHE) sky. H.E.S.S. primarily observes the gamma-ray sky above 100 GeV, where PWNe are more frequently detected than SNRs, with the hardest PWNe being difficult to detect above 50 GeV, since the nebular emission in this band can be faint. In contrast, the best sensitivity of the 4FHL catalog is between 50 and 100 GeV, where the numbers of PWNe and SNRs are more balanced and softer SNRs can be detected that may be missed at TeV energies. Thus, the 4FHL provides a fundamental link between space-based observations with {\it Fermi} and ground-based Cherenkov telescopes.

There are 63 4FHL sources located within the footprint of the H.E.S.S. Galactic Plane Survey. Among these, 39 have been detected at TeV energies and associated with known counterparts, while 24 remain undetected. Note that the highest region of exposure for {\it Fermi} (see Figure \ref{fig:expmap}), near the northern celestial pole, is not visible to H.E.S.S., which primarily covers the southern sky.

Figure \ref{fig:sed_sources} shows the SED for a selection of 4FHL sources with known H.E.S.S. counterparts.

4FHL J0834.4$-$4442e is associated with the PWN Vela-X (HESS J0835$-$455). Vela-X shows distinct spectral behavior across energy bands, with a clear hardening toward higher energies. Recent studies \citep[e.g.][]{lange2025velapulsarpulsarwind} suggest that the hard TeV component is produced by high-energy electrons, while the softer MeV-GeV component may arise from a different electron population or from the interaction of the PWN with the surrounding SNR shell.

4FHL J0851.9$-$4620e is associated with the shell-type SNR RX J0852.0$-$4622 (Vela Junior), one of the brightest steady $\gamma$-ray sources above 1 TeV. The combined {\it Fermi}-LAT and H.E.S.S. spectra are well described by a power law with an exponential cutoff at $E_{\rm cut} \simeq 6.7$ TeV, with no significant spectral variation across the remnant \citep{HESS_RXJ0852_2018}. The origin of the emission remains debated, with leptonic, hadronic, and mixed lepto-hadronic scenarios all proposed to explain the GeV-TeV emission \citep[e.g.][]{Lee_2013,Fukui_2017,sharma2023multiwavelengthanalysisgalacticsupernova}. In this context, the 4FHL data provide continuous GeV-TeV coverage that promises to help to further constrain these models.

4FHL J1514.2$-$5909e is associated with the PWN of MSH~15$-$52 (HESS J1514$-$591). The MSH 15$-$52 system comprises the extended PWN emission, characterized above 50 GeV by a hard spectrum ($\Gamma = 1.99 \pm 0.05$), powered by the young pulsar PSR B1509$-$58. In addition, a fainter point-like source (4FHL J1514.1$-$5910) is detected within the extended PWN emission.

4FHL J1824.4$-$1350e is associated with HESS J1825$-$137, a bright and extended PWN powered by the young pulsar PSR J1826$-$1334. The nebula spans roughly $1^{\circ}$ and shows spatially resolved spectral softening with distance from the pulsar, consistent with synchrotron cooling of electrons as they propagate outward \citep{hess1826,Duvidovich_2019}.

As mentioned before, the improved sensitivity of 4FHL above 50 GeV enables the detection of fainter Galactic sources that were not detected in earlier hard-source catalogs and remain below the current IACT sensitivity. 

4FHL J1124.9$-$5915 ($\Gamma = 1.41 \pm 0.52$) is classified as spp in 4FHL and is spatially coincident with the young oxygen-rich remnant G292.0$+$1.8 \citep{Gaensler2003,Temim2022}. Its very hard spectrum above 50 GeV favors its origin in the compact PWN powered by PSR J1124$-$5916 rather than in the SNR shell. The SED in Figure~\ref{fig:sed_sources} shows the cut-off 4FGL spectrum at a few GeV, dominated by the pulsar emission, while the 4FHL spectrum above 50 GeV is consistent with high-energy emission from the PWN.

Regarding sources detected by H.E.S.S. but missing in 4FHL, we find that most are associated with PWNe (e.g. HESS J1813$-$126, HESS J1846$-$029, HESS J1848$-$018). We also note that, for several H.E.S.S. detections without a nearby 4FHL counterpart, the closest 4FGL source is flagged as confused (e.g. HESS J1626$-$490, HESS J1848$-$018, HESS J1852$-$000).

Thus, the 4FHL catalog and the H.E.S.S. Galactic Plane Survey provide complementary views of the VHE sky in our Galaxy, further supported by other TeV observatories such as VERITAS \citep{VERITAS}, HAWC \citep{HAWC}, and LHAASO \citep{LHAASO}. 

\subsubsection{Comparison with other TeV observatories}

Within the low-latitude VERITAS population, we find 16 of 30 sources with a 4FHL counterpart. Among the 14 sources reported in VERITAS but not in 4FHL, more than 40\% are classified as PSR/PWN systems. This is also consistent with the MAGIC Galactic-plane detections, where 21 of 31 sources have a 4FHL counterpart, while most of the remaining sources are associated with PSR/PWN. VERITAS has carried out deep observations of the Cygnus region \citep{VeritasCygnus}, one of the most active regions of our Galaxy with many sources of GeV and TeV $\gamma$-ray emission. In this region, three 4FHL sources are detected: the PWN 4FHL J2016.1+3712 (VER J2016+371), the Gamma Cygni SNR 4FHL J2020.9+4030e (VER J2019+407), and the SFR 4FHL J2028.6+4110e (MGRO J2031+41). 

Moreover, ground-based large FOV instruments offer an essential complementary view of the TeV $\gamma$-ray and cosmic-ray sky. HAWC, a water Cherenkov observatory, has been detecting TeV sources since 2015. Its latest release \citep{3HWC} reports 65 sources in the northern hemisphere, mostly associated with pulsars and their wind nebulae, and introduces the new class of pulsar halos. In 4FHL, only the Vela pulsar is detected above 50 GeV (4FHL J0835.4$-$4511). Because pulsar emission typically dominates up to $\sim$10 GeV and shows a cutoff at higher energies, it is generally more likely to detect the PWN above 50 GeV. There are five Galactic 4FHL sources within $\sim0.2^{\circ}$ of a HAWC detection, including two PWNe (the Crab Nebula and HESS J1857+026) and systems associated with SNRs, such as IC 443. 

Similarly, LHAASO has recently reported 90 VHE sources \citep{1LHAASO} at declinations between $-20^{\circ}$ and $+80^{\circ}$, with 75 sources detected above 25 TeV, mostly associated with pulsars or PWNe. We find 16 4FHL sources within $0.2^\circ$ of a LHAASO counterpart, including two sources that are reported as new in this work (see section \ref{sec:new_gal}).

\subsubsection{New low-latitude detections with {\it Fermi}} \label{sec:new_gal}
Among the 20 sources without an associated {\it Fermi}-LAT counterpart, five lie at low Galactic latitudes.
Their positions and spectral properties are listed in Table \ref{tab:4fhl_not_in_fermi}.

4FHL J0432.0+5603 ($\Gamma = 1.74 \pm 0.44$) is detected in the northern lobe of SNR G150.3+04.5/1LHAASO J0428$+$5331 (see end of \S~\ref{sec:extended}). Its spectrum is harder than that of the SNR, suggesting the possibility that we are detecting a hotspot inside the remnant or perhaps a previously unidentified source.

4FHL J0501.2+4432 ($\Gamma = 1.89 \pm 0.48$) is spatially coincident with the extended TeV source 1LHAASO J0500$+$4454. Proposed interpretations include a PWN powered by the magnetar SGR 0501$+$4516, cosmic-ray interactions with a molecular cloud, or an SNR \citep{Alford_2026}.

4FHL J1213.2$-$6253 ($\Gamma = 1.95 \pm 0.62$) lies within the extended emission region of 1FGES J1213.3$-$6240, which was included as an extended source in our analysis but not detected. It is associated with the SNR candidate G298.5$-$0.3. However, recent radio observations \citep{SARAOSNR} show that G298.5$-$0.3 shows a morphology inconsistent with known SNRs, and it has been classified as an ‘unusual’ source.

\begin{figure*}[htbp]
  \centering
  \includegraphics[width=1.5\textwidth, trim=400 0 0 0, clip]{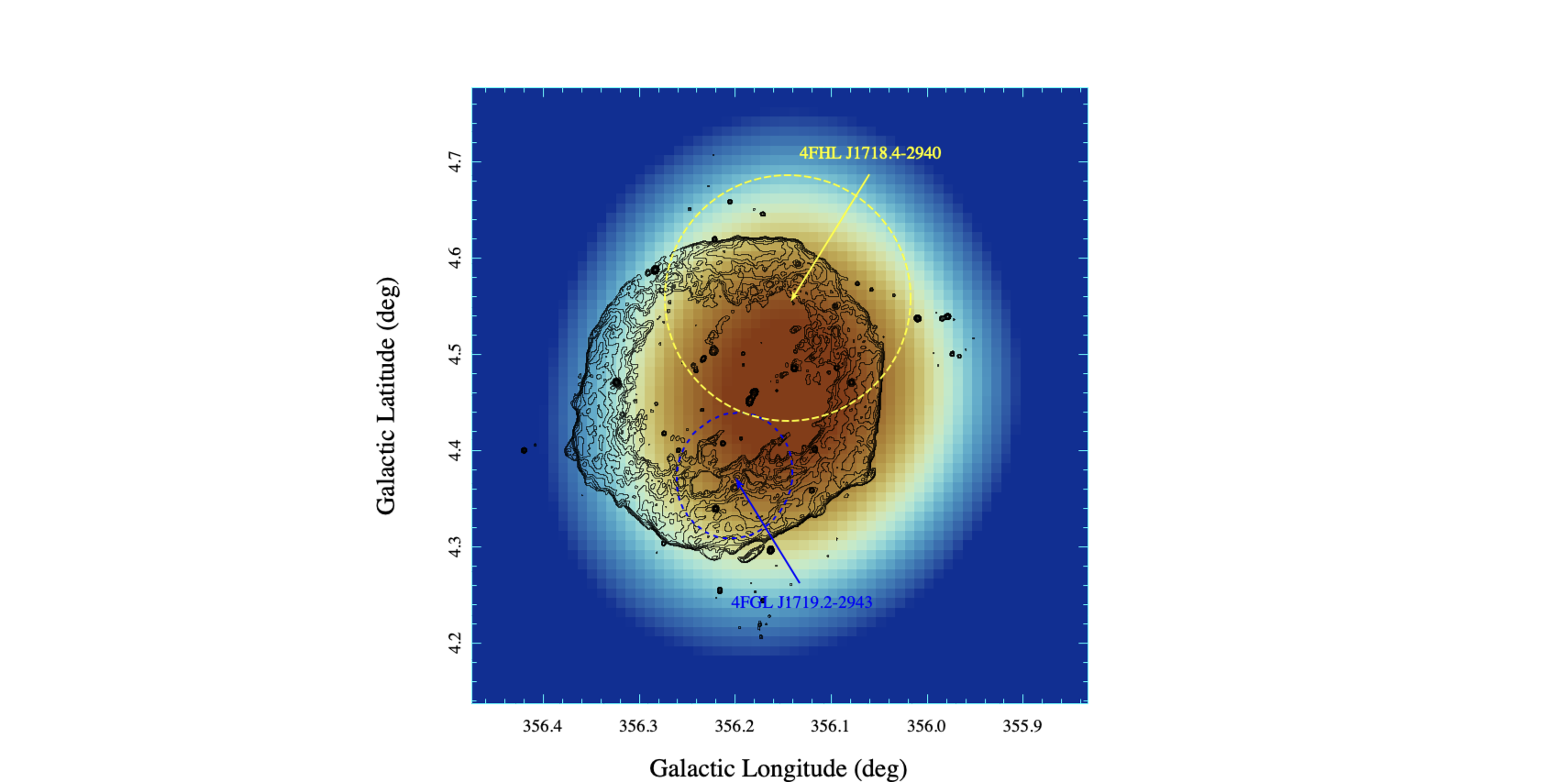}
  \caption{Adaptively smoothed {\it Fermi}-LAT count map above 50 GeV in Galactic coordinates, centered on the region surrounding 4FHL J1718.4$-$2940. Both the 4FHL and 4FGL detections are marked, together with the 1.3 GHz radio contours (black) from the MeerKAT 1.3 GHz observations of SNRs \citep{Cotton_2024}.}
  \label{fig:radio}
\end{figure*}

4FHL J1718.4$-$2940 ($\Gamma = 2.12 \pm 0.46$) is associated with the shell-type SNR G356.2$+$4.5, which shows a well-defined circular morphology in radio observations \citep{Duncan1997}. The 4FGL source (4FGL J1719.2$-$2943) associated with SNR G356.2$+$4.5 lies about 0.2$^{\circ}$ from the 4FHL detection, and the two positions do not overlap when considering their 95\% positional uncertainties. This suggests that the 4FGL and 4FHL detections may be tracing different regions of the SNR. Figure \ref{fig:radio} shows the count map above 50 GeV, where both LAT detections are marked together with the radio contours from the 1.3 GHz MeerKAT  observations \citep{Cotton_2024}.

4FHL J1729.9$-$3336 ($\Gamma = 1.95 \pm 0.45$) is a hard source not previously detected at TeV energies. A bright radio source, NVSS J172951$-$333616 \citep{NVSS}, lies within its small localization area ($r_{95}=0.033^\circ$, $\sim 2$ arcmin) and has a 1.4 GHz flux density of $174.9 \pm 7.3$ mJy, although its nature remains unknown. This source may therefore be a good candidate for a new young SNR or a PWN powered by a young pulsar.

\subsection{Extragalactic \& High-Latitude Population} \label{sec:extragalactic}

A large fraction of 4FHL sources (84.5\%) are either associated with extragalactic sources or located at high latitudes ($|b|\geq 10^{\circ}$). Extragalactic sources tend to show softer observed spectra, with a median observed photon index of $\Gamma \sim 3.05$ ($\Gamma \sim 3.14$ in 2FHL), and are fainter, with a median flux of $F_{50} \simeq 9.3 \times 10^{-12}\ \mathrm{ph\ cm^{-2}\ s^{-1}}$ ($1.85 \times 10^{-11}\ \mathrm{ph\ cm^{-2}\ s^{-1}}$ in 2FHL).

Most of the blazars detected above 50 GeV are of the BL Lacertae object (BL Lac) type, which represents 75\% of the entire 4FHL catalog and 88\% of the extragalactic sample. The majority are High Synchrotron Peak (HSP) blazars, with 70\% of the BL Lacs included in the 3HSP catalog. In contrast, the Low (LSP) and Intermediate-Synchrotron Peak (ISP) types are more commonly detected in lower-energy catalogs such as 4FGL-DR4. Additionally, we detect all HSP blazars reported above 100 GeV in \cite{Agarwal_2025}, with the exception of 4FGL J1807.2+6429, whose test statistic is TS = 25 in  \cite{Agarwal_2025} and TS = 24 in ours.
Only 29 Flat-Spectrum Radio Quasars (FSRQs) are detected, consistent with their typically soft observed spectra. However, this represents an increase of more than 65\% compared to the number of FSRQs detected in 2FHL. The remaining extragalactic sources include 25 blazars of uncertain type (BCUs), one Seyfert 1 active galaxy (PKS 0521$-$365), 11 radio galaxies (RDGs), and 16 remain unidentified. Regarding the high latitude Galactic sources, we detect one Low-mass X-ray binary (LMB; PSR J1023+0038), one SNR (SN 1006), two PWNe (N 157B, CTA 1) and one SPP (SNR G119.5+10.2).

There are 264 4FHL sources in the extragalactic sample that are also detected in 4FGL-DR4, 2FHL, and 3FHL. In contrast, there are 26 other sources that are neither in 4FGL, nor in 2FHL/3FHL, with 15 having no associated LAT counterpart (see \S ~\ref{sec:new_high-lat} for further details). These sources show a power-law index distribution similar to that of the previously detected extragalactic population (median $\Gamma \sim 3.1$ vs. $\Gamma \sim 3.0$), but they are systematically fainter (median $F_{50} \sim 7.7 \times 10^{-12}$ ph cm$^{-2}$ s$^{-1} $ vs $F_{50} \sim 1.72 \times 10^{-11}$ ph cm$^{-2}$ s$^{-1} $) than those detected in the previously mentioned catalogs. Relative to the VHE active galactic nuclei (AGN) catalog above 100 GeV \citep{neronov2025catalogveryhighenergyemittingactive}, which contains 275 sources (dominated by BL Lacs) and only seven FSRQs, 4FHL detects nearly 80\% of the VHE sources reported while substantially increasing the number of identified FSRQs. Additionally, we detect all high-latitude photon clusters listed in \cite{Pshirkov_2025}.

Redshifts are available in the literature \citep[e.g.][]{4LAC,Foschini_2022,1CGRH,Firmamento} for 433 sources ($\sim 76\%$ of the 4FHL AGN population), of which 127 lack redshift information in the latest release of the \textit{Fermi}-LAT AGN catalog \citep[4LAC-DR3;][]{4LACDR3}. The redshift distribution has a median value of $z \sim 0.3$, higher than that reported in the 2FHL catalog ($z \sim 0.2$), reflecting the inclusion of fainter and more distant sources in 4FHL.  
While BL Lacs generally lie at lower redshifts than FSRQs, the limited number of FSRQs detected above 50 GeV results in the high-redshift population of the 4FHL being dominated by BL Lacs. In total, 36 BL Lacs with $z>1$ are reported, including four at $z>2$. 4FHL J1006.1+6440 is the highest-redshift BL Lac reported in the catalog, with $z\sim4$. We note, however, that this value comes from SDSS \citep{SDSS} and should be treated with caution, as it appears to rely on a photometric estimate and the optical spectrum is essentially featureless. Moreover, the 4FHL detection above 50 GeV and the observed photon index of $\Gamma=2.44\pm0.95$ are not easily reconciled with a redshift of $z\sim4$, where standard EBL models predict strong attenuation.

The 4FHL catalog also provides an important resource for EBL studies and its effect on gamma-ray propagation \citep[e.g.][]{Dwek_2013,Dominguez_2015}. At energies above tens of GeV, gamma rays from distant blazars are attenuated through pair production with EBL photons, an effect that imprints a redshift- and energy-dependent softening in their observed spectra \citep[e.g.][]{Gould_1967,Ackermann_2012,EBL_HESS,EBL_Acciari_2019,EBL_Abeysekara_2019,EBL_Greaux_2024}. The number of high-latitude and redshift (z$>1$) blazars detected above 50 GeV enables population-level analyses of EBL-induced absorption and tests of the cosmic gamma-ray horizon \citep[CGRH;][]{Dominguez_2013,1CGRH}.

\subsubsection{Comparison with TeV Extragalactic Surveys}\label{sec:HEGS}

The H.E.S.S. collaboration has conducted observations of the extragalactic sky in VHE (\(E > 100 \, \text{GeV}\)) covering 5.7\% of the sky at \(|b| \geq 10^{\circ}\), resulting in the first H.E.S.S. Extragalactic Survey \citep[HEGS;][]{HEGS_2025}.  

The survey led to the detection of 23 extragalactic sources, all of which were already established as VHE emitters, including 20 AGN (predominantly HSP BL Lacs), two radio galaxies, and the starburst galaxy NGC~253. One additional source, PKS~0625$-$354, remains of uncertain type. All of these sources are also detected in the 4FHL catalog, with the exception of NGC~253, which has a test statistic of TS $=17$ in our analysis and therefore is not reported in the final catalog. Consistent with the 4FHL results, this suggests that HSP BL Lacs dominate the resolved VHE extragalactic source population and provide an important contribution to the extragalactic $\gamma$-ray background (EGB).

\begin{figure}[htbp]
    \centering
    \begin{minipage}{0.45\textwidth}
        \centering
        \includegraphics[width=\linewidth]{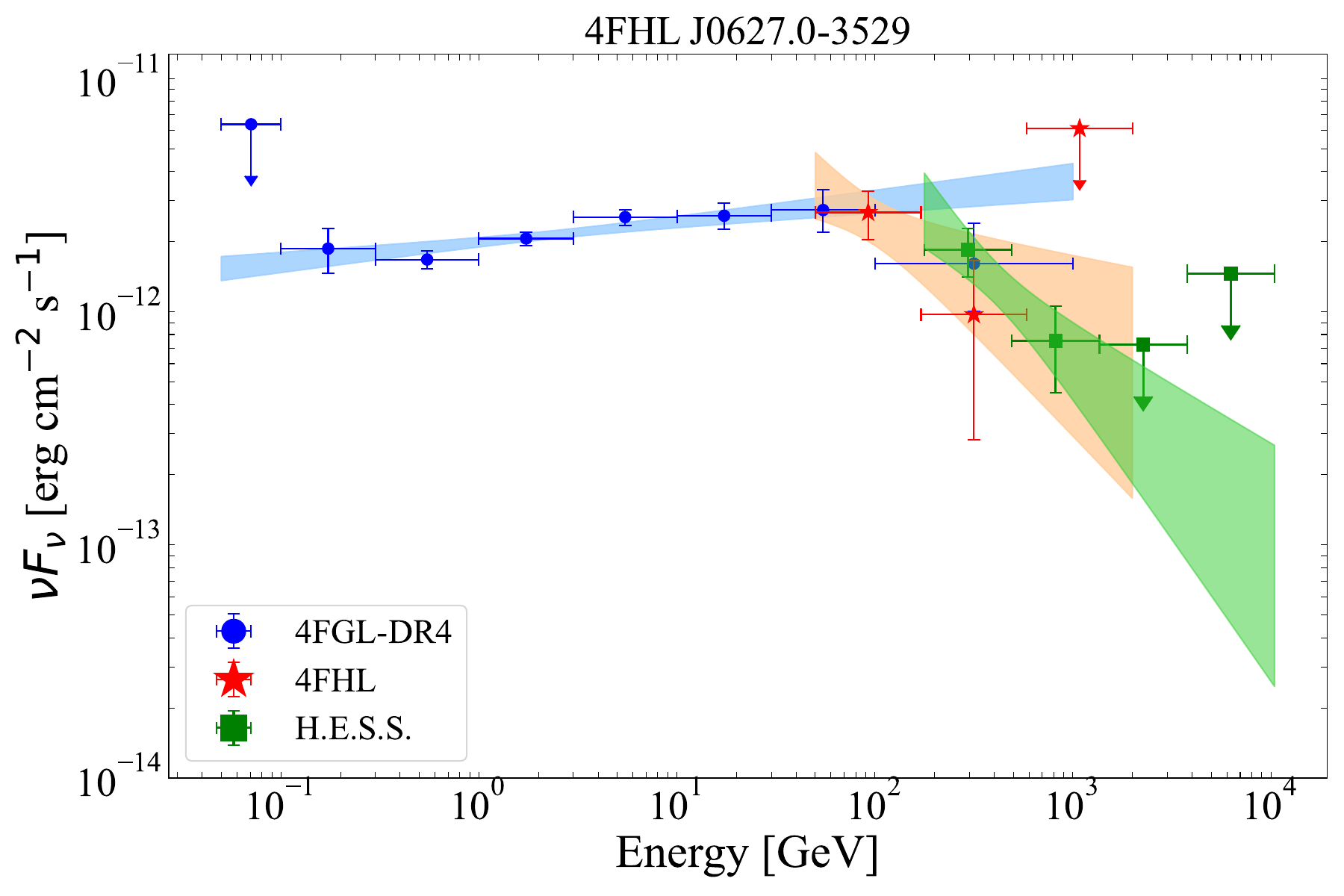} 
    \end{minipage}
    \hspace{0.05\textwidth} 
    \begin{minipage}{0.45\textwidth}
        \centering
        \includegraphics[width=\linewidth]{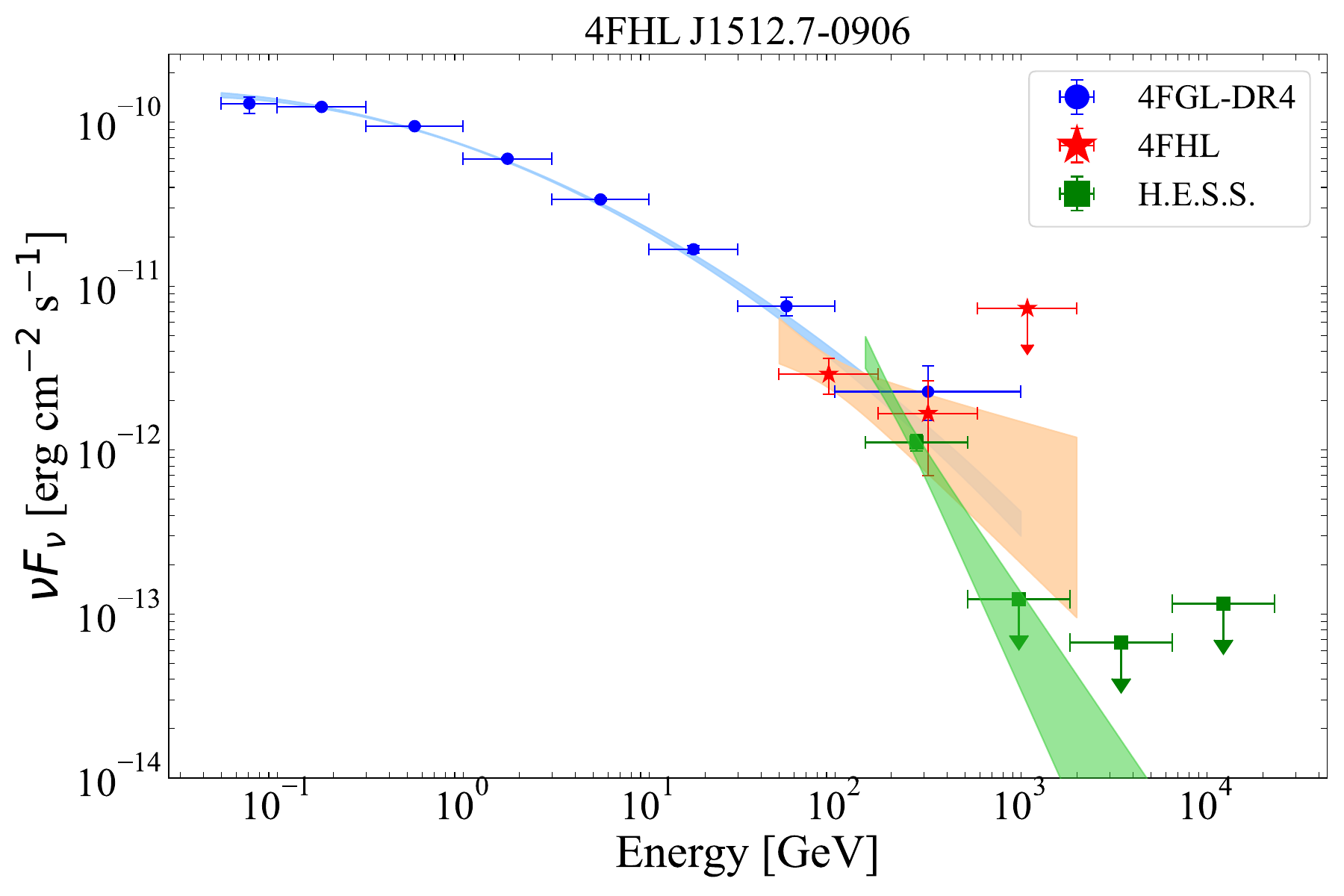}
    \end{minipage}

    \begin{minipage}{0.45\textwidth}
        \centering
        \includegraphics[width=\linewidth]{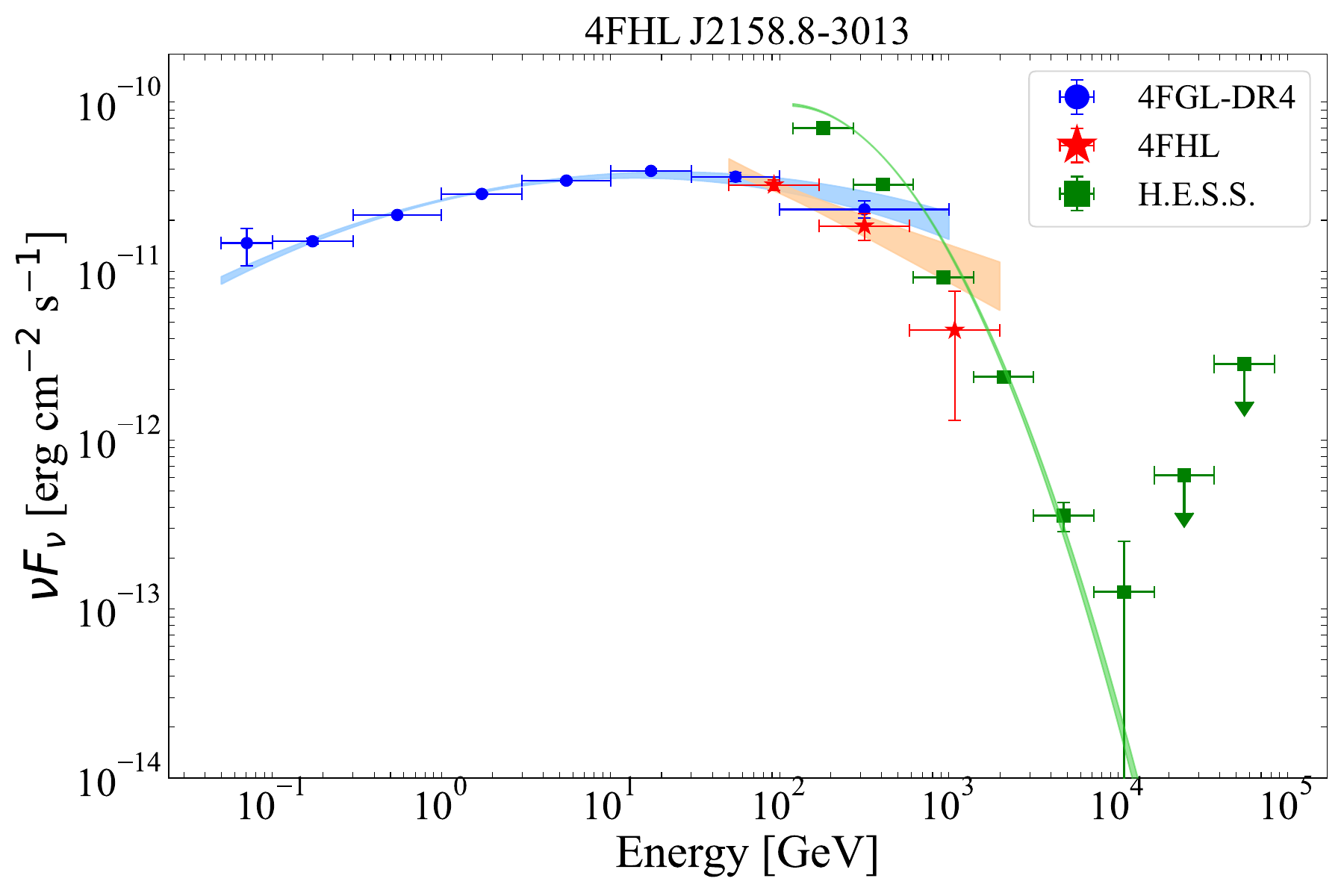} 
    \end{minipage}
    \hspace{0.05\textwidth} 
    \begin{minipage}{0.45\textwidth}
        \centering
        \includegraphics[width=\linewidth]{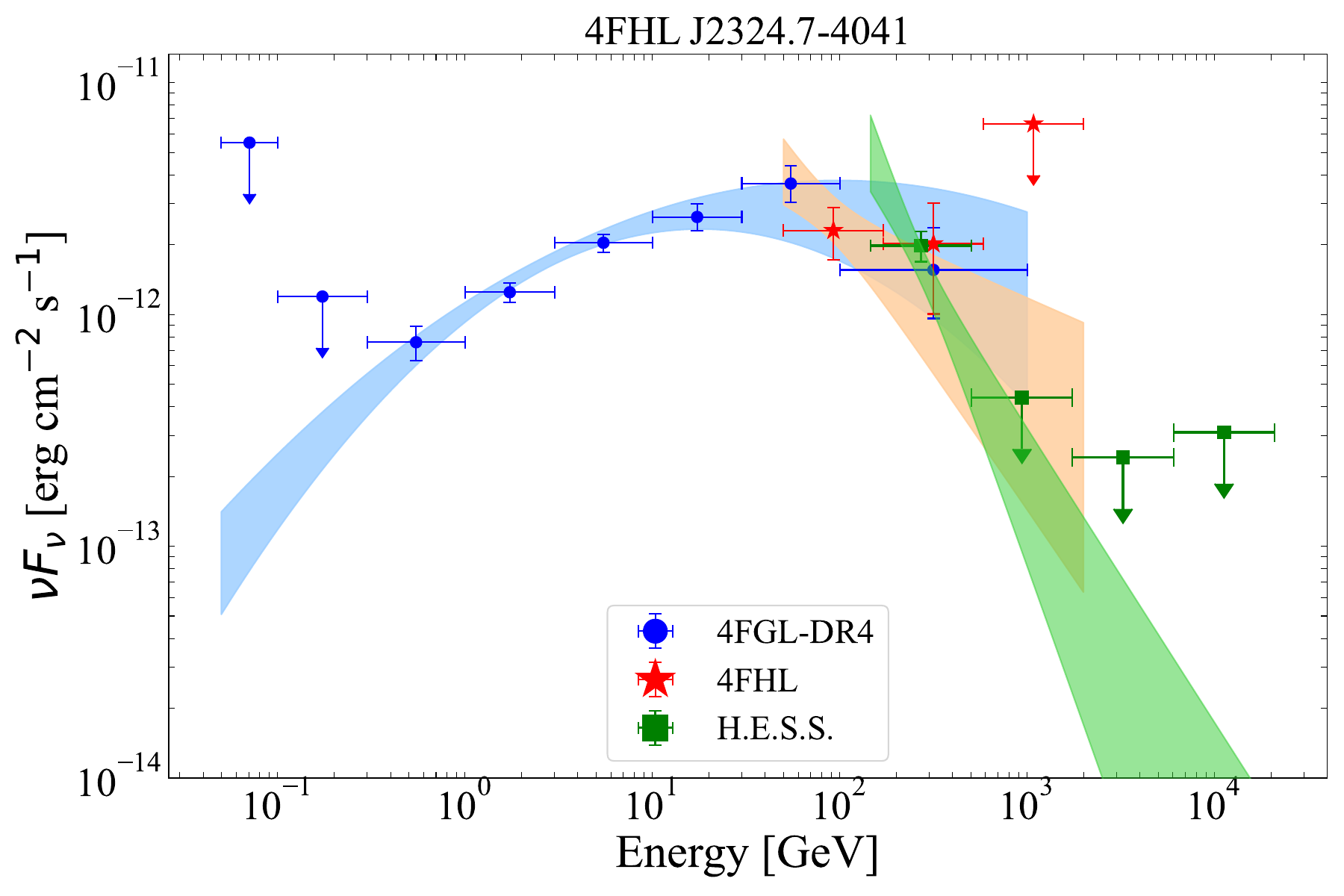} 
    \end{minipage}

    \caption{Spectral energy distributions of four extragalactic sources discussed in \S \ref{sec:HEGS}: (top left)  Radio galaxy PKS~0625$-$35, (top right) FSRQ PKS 1510$-$089, (bottom left) HSP blazar PKS 2155$-$304, and the HSP blazar 1ES 2322$-$409 (bottom right). The data include 4FGL-DR4 (blue dots), 4FHL (red stars),  and H.E.S.S. (green squares).}
    \label{fig:sed_sources_extra}
\end{figure}

Figure \ref{fig:sed_sources_extra} shows the SED for sources of the three different extragalactic populations detected by H.E.S.S. for which the LAT now resolves the descending contribution of the high energy peak.

4FHL J0627.0$-$3529 is associated with the radio galaxy PKS 0625$-$35. 
The broadband $\gamma$-ray spectrum shows a clear softening at high energies. While it is often modeled with a simple power law, several studies \citep[e.g.][]{Baghmanyan_2018,Sahakyan_2018} have shown that a power law with an exponential cutoff provides a better description, with curvature emerging around $\sim 100$ GeV. The inclusion of 4FHL data strengthens this picture, as the combined {\it Fermi}-LAT (4FGL and 4FHL) and H.E.S.S. SED makes the curvature more evident.

4FHL J1512.7$-$0906 is associated with the very active FSRQ PKS 1510$-$089. In the 4FHL catalog, it represents the hardest  FSRQ detected ($\Gamma = 2.16 \pm 0.50$). Multiwavelength modeling indicates that the high-energy component requires two distinct IC processes: seed photons from the broad line region account for the GeV emission observed by {\it Fermi}-LAT, whereas the TeV emission is more naturally explained by IC scattering of photons from the dusty torus, a scenario that also reproduces the X-ray band \citep{Kataoka_2008,Barnacka_2014}. Within this framework, a leptonic origin provides a consistent explanation for the broadband emission, although more complex scenarios involving multiple external radiation fields have also been proposed to capture the variability and flare behavior observed across different epochs \citep{marscher2010innerjetquasarpks,lei2025insightoriginmultiwavelengthemissions}.

4FHL J2158.8$-$3013 is associated with the HSP blazar PKS 2155$-$304, one of the best-studied blazars due to its extreme variability across all wavelengths and on multiple timescales, often showing energy-dependent behavior \citep{harutyunyan2025comprehensiveviewpks2155304}. The source shows a spectral softening toward higher energies in combination with spectral variability on its emission. Its spectrum typically hardens during flux-rise phases when particle acceleration dominates over cooling, and softening in the opposite case \citep{Kirk_1997}. The high-energy emission is well described by synchrotron self-Compton (SSC) scattering \citep{Tavecchio_1998}, where the same population of electrons upscatters synchrotron photons into the GeV$-$TeV band. However, alternative scenarios involving external Compton or hadronic processes have also been suggested \citep{Barkov_2012,Gao_2021}.

4FHL J2324.7$-$4041 is associated with the HSP blazar 1ES 2322$-$409. Similar to PKS 2155$-$304, this source shows a spectral softening toward higher energies. The broadband emission is well described by the SSC model \citep{Abdalla_2019}. Unlike other variable HSP blazars, no significant evidence of high-energy variability has been reported, which would further support the SSC scenario as the dominant emission mechanism \citep{Goswami_2023}.

Relative to ground-based TeV observations, we detect all VERITAS extragalactic sources except the starburst galaxy M82 and the FSRQ 3C 273 \citep{M82_Veritas,VERITAS_3c273}, consistent with the 4FHL extragalactic population being dominated by BL Lacs (HSP).  This is also consistent with the MAGIC extragalactic sample, for which all sources are included in 4FHL except the BL Lac GB6 J1058$+$2817, which was detected by MAGIC during a flaring state with a soft VHE spectrum \citep{MAGIC_GB6,Cerasole_2025}. In addition, apart from the well-known Markarian blazars \citep{mrk421,mrk501}, LHAASO \citep{1LHAASO} has reported several HSP blazars that are also detected in 4FHL, including 4FHL J1728.3+5013, 4FHL J2026.8+3340, and 4FHL J2347.0+5141.

\subsubsection{New high-latitude detections with {\it Fermi}}\label{sec:new_high-lat}

We report 15 sources located at $|b|\geq 10^{\circ}$ that were not detected in previous LAT catalogs. These sources are mostly faint and eight remain unassociated. Table \ref{tab:4fhl_not_in_fermi} shows the spectral properties of these sources along with their closest 4FGL-DR4 detection. Among them, six are classified as blazars (four HSP blazars, one BCU and one FSRQ PKS 1351+021). There are seven sources with nearby ($\leq0.2^{\circ}$) 4FGL counterparts that were not associated through the association procedure (see Sect.~\ref{sec:association}). These cases may represent the same sources or cases of source confusion.

4FHL J0810.9+3533 ($\Gamma = 0.69 \pm 0.59$) is a very hard and faint source at high Galactic latitude ($b \simeq 31^{\circ}$), with no nearby 4FGL source. Within its localization region ($r_{95}=0.171^\circ$, $\sim 10.3$ arcmin), the brightest known radio source is GB2 0806+356 \citep{Machalski1998}, with a 1.4 GHz flux density of $260 \pm 6$ mJy and photon index $\Gamma_{1.4 \rm GHz} = 0.65 \pm 0.02$. 

4FHL J1120.5$-$2648 ($\Gamma = 3.50 \pm 0.93$) shows a soft spectrum typical of blazars. It is associated with the BCU NVSS J112031$-$264828. The closest 4FGL counterpart (0.09$^{\circ}$) is the unassociated source 4FGL J1120.1$-$2645.

4FHL J1128.2$-$4919 ($\Gamma = 2.98 \pm 1.02$) is an unassociated source with a spectral index consistent with a blazar. The nearest 4FGL source, 4FGL~J1127.6$-$4920 (0.1$^{\circ}$), is the radio source MRC~1125$-$490, classified as a BCU. Their similar spectral properties and relatively small angular separation suggest that they could be the same source.

4FHL J1314.8$+$2358 is a faint, unassociated source with limited photon statistics ($N_{\mathrm{pred}}\simeq 3.5$) and likely a soft spectrum. It lies $0.2^{\circ}$ from the BL Lac 4FGL J1314.7$+$2348 (TXS 1312$+$240). Similarly, 4FHL J1339.4$+$1150 ($\Gamma = 2.90 \pm 0.99$) lies $0.15^{\circ}$ from the BL Lac 4FGL J1338.9$+$1153 (SDSS J133859.05$+$115316.7), which has a harder spectrum in 4FGL.

4FHL J1503.7$-$4141 ($\Gamma = 0.99 \pm 0.41$) is associated with SN 1006 (SNR G327.6$+$14.6). 4FGL J1503.6$-$4146 is also associated with SN 1006, lying $\sim$0.1$^{\circ}$ north-east of the 4FHL position, suggesting that both detections may trace different regions of the remnant. The spectrum in 4FGL is softer than in 4FHL, consistent with emission from zones of different ambient density: hard gamma-ray emission arising from low-density regions dominated by IC scattering, and softer emission from higher-density regions where hadronic interactions contribute \citep[e.g.][]{Yuan2012, Zeng_2019}. The very hard 4FHL spectrum supports efficient particle acceleration in the remnant \citep{Giuffrida_2022,Lemoine_Goumard_2025}.

4FHL J1605.2$+$5420 and 4FHL J2024.6$-$0847 are both associated with HSP BL Lacs, RBS 1555 and 1RXS J202428.9$-$084810, and have close 4FGL detections, 4FGL J1605.5$+$5423 (0.07$^{\circ}$) and 4FGL J2024.4$-$0847 (0.06$^{\circ}$), which are associated with the same blazars. They might therefore correspond to the same source.

\subsection{Candidates for detection with Cherenkov Telescopes}
\begin{figure}[htbp]
\centering
\includegraphics[width=1\textwidth]{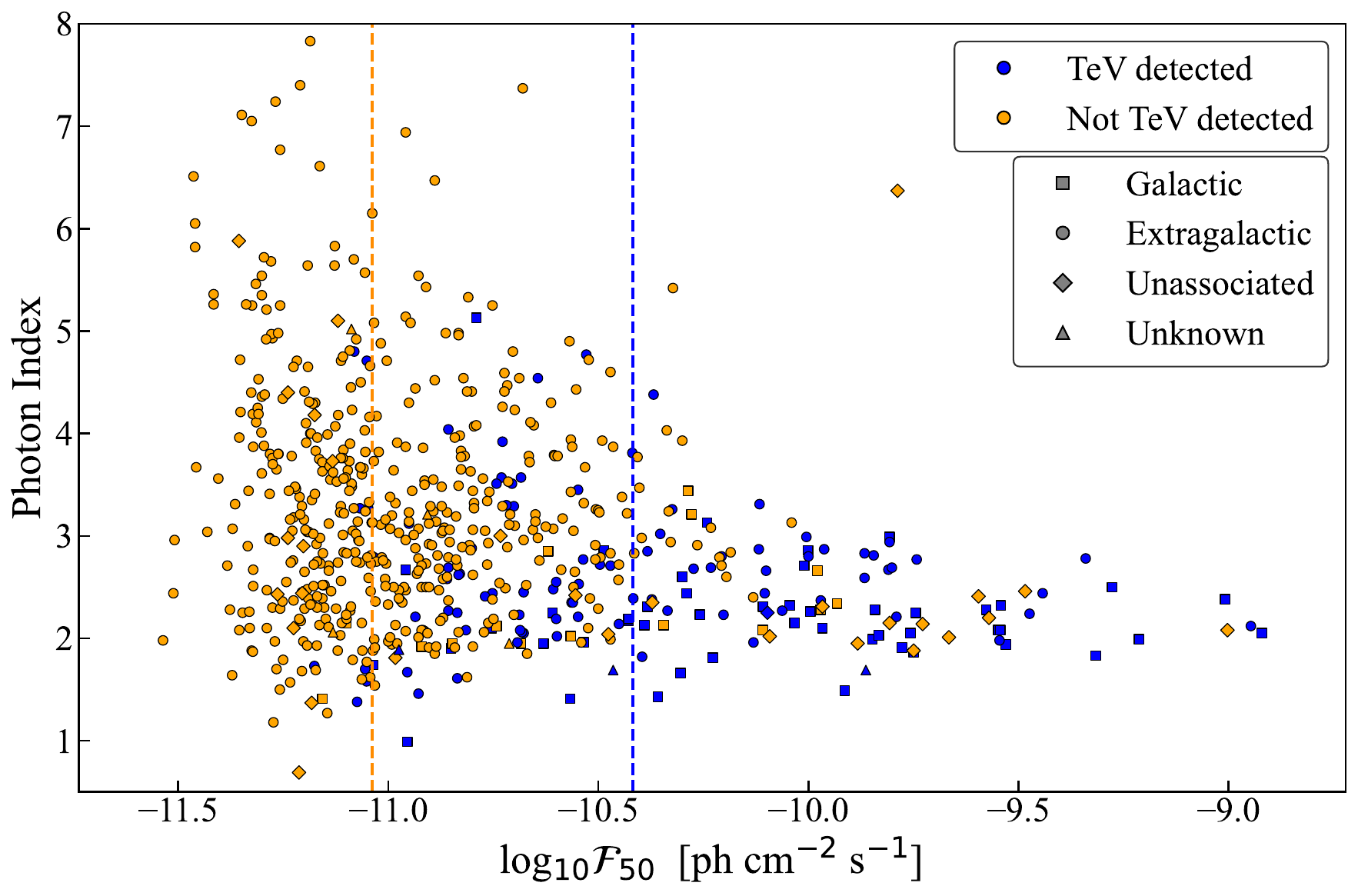} 
\caption{Observed photon index as a function of the integrated photon flux is shown for sources detected in the 4FHL catalog. Galactic sources are indicated by squares, extragalactic sources by circles, unassociated sources by diamonds, and sources of unknown type by triangles. Blue symbols represent sources detected by IACTs, while orange symbols correspond to the $\sim$80\% of the 4FHL sample that remain undetected at TeV energies. The median flux values for sources detected and not detected by IACTs are indicated by colored vertical lines. }
\label{fig:iIACTcandidates}
\end{figure}

Imaging Atmospheric Cherenkov Telescopes (IACTs) provide excellent sensitivity in the TeV domain but are limited by their narrow field of view. A cross-correlation between the 4FHL and TeVCat catalogs shows that 529 sources, i.e. $\sim 79\%$ of the catalog, have no TeVCat counterpart. Considering the different field of view, duty cycle, sensitivity, exposure, and observing strategy of LAT and IACTs, these sources provide a useful reference sample for guiding future TeV observations. Figure~\ref{fig:iIACTcandidates} presents the photon index and flux for sources detected and not detected by IACTs. On average, the flux of sources already detected by IACTs is about four times higher than that of the non-detected population. Among the 529 non-detected sources at TeV energies, 94 are located at $|b| < 10^{\circ}$, while 435 (i.e. $\sim 82\%$) are at higher latitudes ($|b| \geq 10^{\circ}$), indicating that most 4FHL candidates for future IACT observations are likely extragalactic. Notably, 25 sources not detected by IACTs have significant emission in the 171-585 GeV band,  making them good candidates for future follow-up observations.

However, because of the improved sensitivity of 4FHL, the majority of these $\gamma$-ray sources are relatively faint and may remain below the reach of current-generation IACTs such as H.E.S.S., VERITAS, or MAGIC, which achieve sensitivities of $1$-$2\%$ of the Crab Nebula flux above 100 GeV for 50 hours of observation. The upcoming Cherenkov Telescope Array Observatory \citep[CTAO;][]{CTAO}, with a projected sensitivity up to an order of magnitude better\footnote{For current IACTs, we assume an average sensitivity of $\sim$3-4\% of the Crab Nebula flux  ($10^{-9}\ \mathrm{ph\ cm^{-2}\ s^{-1}}$) in the 50 GeV-2 TeV range, since the Crab spectrum is softer above 1 TeV.}, will be able to detect a significant fraction of the sources, making the 4FHL catalog an important resource for planning future IACT observations.  

In addition, Figure~\ref{fig:iIACTcandidates} shows promising candidates characterized by both hard photon indices and relatively large fluxes, including some Galactic sources that are strong, a priori, targets for IACT observations. Within the extragalactic population, about $78\%$ of the high-latitude sources not yet detected by IACTs are associated with HSP blazars, which are particularly favorable for detection since their high-energy peak typically occurs above 100 GeV.

\input{tables/Table_new}

\section{Summary}\label{sec:summary}
We have presented the Fourth Catalog of Hard \textit{Fermi}-LAT Sources (4FHL), based on 16 years of LAT observations above 50 GeV.  
The 4FHL achieves better sensitivity, localization, and power-law index characterization than previous hard source catalogs, leading to the detection of 673 sources, a factor $\sim$ 2 times more sources than 2FHL in the 50\,GeV$-$2\,TeV range. This catalog represents the most comprehensive LAT view of the high-energy gamma-ray sky above 50 GeV and provides an essential link between the GeV regime probed by broad-band LAT catalogs and the TeV domain explored by ground-based instruments.

The 4FHL source population is dominated by extragalactic sources, which represent about 84\% of the detections.  Most of these are blazars, with BL Lacs being the largest class (75\% of the entire catalog and 88\% of the AGN sample).  Most BL Lacs in 4FHL belong to the HSP class, which display the hardest $\gamma$-ray spectrum (among blazars) and remain bright in the 50\,GeV$-$2\,TeV band, making them essential targets for EGB and EBL studies. FSRQs are more difficult to detect due to their soft spectra; nevertheless, 29 are included in 4FHL, representing an increase of more than 65\% compared to 2FHL.  
In addition, we report 25 blazars of uncertain type, one Seyfert 1 galaxy (PKS~0521$-$365), and 11 radio galaxies.

Within the Galactic plane, the 4FHL includes 156 sources ($|b|<10^{\circ}$). While a fraction of sources are extragalactic ($\sim$ 47\%), the rest are Galactic or unassociated.  The Galactic  population is characterized by hard photon indices (median $\Gamma \sim 2.1$), reflecting efficient particle acceleration up to TeV energies. Most of them are associated with late stages of stellar evolution, such as PWNe and SNRs. Additionally, six binary systems, one pulsar, and two SFRs are also detected. Notably, 55 extended sources are detected, indicating that more than half of the LAT extended population are significant above 50 GeV.

Only a small fraction of sources ($\sim$5\%) remain unassociated. Several of these objects are faint, hard, and are distributed across both low and high Galactic latitudes.  
In total, 20 lack an associated counterpart with respect to all previously detected LAT sources.

The 4FHL catalog thus provides an extensive overview of the gamma-ray sky above 50 GeV, serving as a valuable resource for population studies, future multiwavelength studies, and follow-up observations with current and future ground-based $\gamma$-ray observatories.


\section*{Acknowledgments}
The \textit{Fermi}-LAT Collaboration acknowledges generous ongoing support
from a number of agencies and institutes that have supported both the
development and the operation of the LAT as well as scientific data analysis.
These include the National Aeronautics and Space Administration and the
Department of Energy in the United States, the Commissariat \`a l'Energie Atomique
and the Centre National de la Recherche Scientifique / Institut National de Physique
Nucl\'eaire et de Physique des Particules in France, the Agenzia Spaziale Italiana
and the Istituto Nazionale di Fisica Nucleare in Italy, the Ministry of Education,
Culture, Sports, Science and Technology (MEXT), High Energy Accelerator Research
Organization (KEK) and Japan Aerospace Exploration Agency (JAXA) in Japan, and
the K.~A.~Wallenberg Foundation, the Swedish Research Council and the
Swedish National Space Board in Sweden.

Additional support for science analysis during the operations phase is gratefully 
acknowledged from the Istituto Nazionale di Astrofisica in Italy and the Centre 
National d'\'Etudes Spatiales in France. This work was performed in part under DOE 
Contract DE-AC02-76SF00515.

This work was supported by the \textit{Fermi} Guest Investigator Cycle 17 program (80NSSC25K7894).

This research has made use of data and/or software provided by the High Energy Astrophysics Science Archive Research Center (HEASARC), which is a service of the Astrophysics Science Division at NASA/GSFC.

\bibliography{bibliography}
\bibliographystyle{aasjournalv7}

\newpage

\newpage

\clearpage

\clearpage

\clearpage

\clearpage

\newpage

\clearpage

\newpage

\clearpage

\end{document}

%% file: authors_orcid.tex
\author[orcid=0000-0002-6606-2816]{F.~Acero}
\email{default@example.com}
\affiliation{Universit\'e Paris-Saclay, Universit\'e Paris Cit\'e, CEA, CNRS, AIM, F-91191 Gif-sur-Yvette Cedex, France}
\affiliation{FSLAC IRL 2009, CNRS/IAC, La Laguna, Tenerife, Spain}
\author[orcid=0009-0004-3923-9884]{A.~Adelfio}
\email{default@example.com}
\affiliation{Istituto Nazionale di Fisica Nucleare, Sezione di Perugia, I-06123 Perugia, Italy}
\author[orcid=0000-0002-6584-1703]{M.~Ajello}
\email[show]{majello@clemson.edu}
\affiliation{Department of Physics and Astronomy, Clemson University, Kinard Lab of Physics, Clemson, SC 29634-0978, USA}
\author[orcid=0009-0001-2927-8968]{E.~Aviano}
\email{default@example.com}
\affiliation{Dipartimento di Fisica, Universit\`a di Trieste, I-34127 Trieste, Italy}
\affiliation{Istituto Nazionale di Fisica Nucleare, Sezione di Trieste, I-34127 Trieste, Italy}
\author[orcid=0000-0002-9785-7726]{L.~Baldini}
\email{default@example.com}
\affiliation{Universit\`a di Pisa, Dipartimento di Fisica E. Fermi, I-56127 Pisa, Italy}
\affiliation{Istituto Nazionale di Fisica Nucleare, Sezione di Pisa, I-56127 Pisa, Italy}
\author[orcid=0000-0002-8784-2977]{J.~Ballet}
\email{default@example.com}
\affiliation{Universit\'e Paris-Saclay, Universit\'e Paris Cit\'e, CEA, CNRS, AIM, F-91191 Gif-sur-Yvette Cedex, France}
\author[orcid=0000-0001-7233-9546]{C.~Bartolini}
\email{default@example.com}
\affiliation{Istituto Nazionale di Fisica Nucleare, Sezione di Bari, I-70126 Bari, Italy}
\affiliation{Universit\`a degli studi di Trento, via Calepina 14, 38122 Trento, Italy}
\author[orcid=0000-0002-6729-9022]{J.~Becerra~Gonzalez}
\email{default@example.com}
\affiliation{Instituto de Astrof\'isica de Canarias and Universidad de La Laguna, Dpto. Astrof\'isica, 38200 La Laguna, Tenerife, Spain}
\author[orcid=0000-0002-2469-7063]{R.~Bellazzini}
\email{default@example.com}
\affiliation{Istituto Nazionale di Fisica Nucleare, Sezione di Pisa, I-56127 Pisa, Italy}
\author[orcid=0000-0001-9935-8106]{E.~Bissaldi}
\email{default@example.com}
\affiliation{Dipartimento di Fisica ``M. Merlin" dell'Universit\`a e del Politecnico di Bari, via Amendola 173, I-70126 Bari, Italy}
\affiliation{Istituto Nazionale di Fisica Nucleare, Sezione di Bari, I-70126 Bari, Italy}
% ...

%% file: tables/Table_class.tex
\begin{deluxetable}{lcrcr}
\setlength{\tabcolsep}{0.04in}
\tablewidth{0pt}
\tabletypesize{\small}
\tablecaption{4FHL Source Classes \label{tab:classes}}
\tablehead{
\colhead{Description} & 
\multicolumn{2}{c}{Identified} &
\multicolumn{2}{c}{Associated} \\
& 
\colhead{Designator} &
\colhead{Number} &
\colhead{Designator} &
\colhead{Number}
}
\startdata
Pulsar & PSR & 1 & psr & \nodata \\
Pulsar Wind Nebula & PWN & 10 & pwn & 6\\
Supernova remnant & SNR & 17 & snr & 4 \\
Supernova remnant / Pulsar wind nebula &  SPP  &  2  & spp & 16 \\
High-mass binary & HMB & 4 & hmb & \nodata \\
Low-mass binary & LMB & 1 & lmb & \nodata \\
Binary & BIN & 1 &  \nodata  &  \nodata \\
Galactic center & GC & 1 & \nodata  &  \nodata \\
Star-forming region & SFR & 2 &  sfr  &  \nodata  \\
\hline
BL Lac type of blazar & BLL & 19 & bll & 485 \\
Flat spectrum radio quasar type of blazar & FSRQ & 8 & fsrq & 21 \\
Non-blazar active galaxy &  AGN  &  1 & agn & \nodata \\
Radio galaxy & RDG & 3 & rdg & 8 \\
Blazar candidate of uncertain type &  BCU  &  1  & bcu & 24 \\
\hline
Total & identified & 71 & associated & 564 \\
\hline
Unclassified &   UNK  &   2 & unk & 5 \\
Unassociated & \nodata & \nodata &  \nodata & 31 \\
Total in 4FHL & \nodata & \nodata & \nodata & 673 \\
\enddata
\tablecomments{ The designation `spp' indicates potential association with SNR or PWN. Designations in capital letters are firm identifications; small letters indicate associations.}
\end{deluxetable}

%% file: tables/Table_description.tex
\begin{deluxetable}{lccl}
\setlength{\tabcolsep}{0.04in}
\tablewidth{0pt}
\tabletypesize{\scriptsize}
\tablecaption{Description of the 4FHL Catalog \label{tab:description4fhl}}
\tablehead{
\colhead{Column} & 
\colhead{Format} &
\colhead{Unit} &
\colhead{Description}
}
\startdata
Source\_Name & 18A & \nodata & 4FHL source name (suffix ``e'' indicates an extended source) \\
RAJ2000 & E & deg & Right Ascension (J2000) \\
DEJ2000 & E & deg & Declination (J2000) \\
GLON & E & deg & Galactic longitude \\
GLAT & E & deg & Galactic latitude \\
Pos\_err\_68 & E & deg &  Radius of position uncertainty circle at 68\% confidence level \\
Pos\_err\_95 & E & deg & Radius of position uncertainty circle at 95\% confidence level\\
TS & E & \nodata & Test Statistic for the full band (50 GeV$-$2 TeV) \\
PL\_Index$^{a}$ & E & \nodata & Observed power–law photon index \\
Unc\_PL\_Index$^{a}$ & E & \nodata & 1$\sigma$ statistical uncertainty on the observed photon index \\
Flux50 & E & ph\,cm$^{-2}$\,s$^{-1}$ & Integrated photon flux (50 GeV$-$2 TeV) \\
Unc\_Flux50 & E & ph\,cm$^{-2}$\,s$^{-1}$ & 1$\sigma$ uncertainty on \texttt{Flux50} \\
Energy\_Flux50 & E & erg\,cm$^{-2}$\,s$^{-1}$ & Energy flux (50 GeV$-$2 TeV) \\
Unc\_Energy\_Flux50 & E & erg\,cm$^{-2}$\,s$^{-1}$ & 1$\sigma$ uncertainty on \texttt{Energy\_Flux50} \\
Flux50\_171GeV & E & ph\,cm$^{-2}$\,s$^{-1}$ & Integrated photon flux from 50 GeV to 171\,GeV \\
Unc\_Flux50\_171GeV$^{b}$ & E & ph\,cm$^{-2}$\,s$^{-1}$ & 1$\sigma$ uncertainty on \texttt{Flux50\_171GeV} \\
Sqrt\_TS50\_171GeV & E & \nodata & $\sqrt{\mathrm{TS}}$ in 50–171\,GeV \\
Flux171\_585GeV & E & ph\,cm$^{-2}$\,s$^{-1}$ & Integrated photon flux from 171 GeV to 585\,GeV \\
Unc\_Flux171\_585GeV$^{b}$ & E & ph\,cm$^{-2}$\,s$^{-1}$ & 1$\sigma$ uncertainty on \texttt{Flux171\_585GeV} \\
Sqrt\_TS171\_585GeV & E & \nodata & $\sqrt{\mathrm{TS}}$ in 171–585\,GeV \\
Flux585\_2000GeV & E & ph\,cm$^{-2}$\,s$^{-1}$ & Integrated photon flux from 585 GeV to 2 TeV \\
Unc\_Flux585\_2000GeV$^{b}$ & E & ph\,cm$^{-2}$\,s$^{-1}$ & 1$\sigma$ uncertainty on \texttt{Flux585\_2000GeV} \\
Sqrt\_TS585\_2000GeV & E & \nodata & $\sqrt{\mathrm{TS}}$ in 585–2000\,GeV \\
Npred & E & \nodata & Predicted number of photons from the source \\
HEP\_Energy & E & GeV & Highest–energy photon associated to the source \\
HEP\_Prob & E & \nodata & Probability that the HEP originates from the source \\
ROI & I & \nodata & Region of interest number\\
ASSOC & 28A & \nodata & Name of the most likely associated source\\
ASSOC\_PROB\_BAY & E & \nodata & Probably of association from the Bayesian method\\
ASSOC\_PROB\_LR & E & \nodata & Probably of association from the likelihood ratio method \\
CLASS & 8A & \nodata & Class designation from the most likely association (see Table \ref{tab:classes}) \\
Redshift & E & \nodata & Redshift (when available) of the  most likely associated source \\
4FGL & 18A & \nodata & 4FGL counterpart name when available \\
2FHL & 18A & \nodata & 2FHL counterpart name when available\\
3FHL & 18A & \nodata & 3FHL counterpart name when available\\
TeV & 28A & \nodata & Name of the most likely associated source in TevCat \\
\enddata
\tablenotetext{a}{ Sources with very large fitted photon indices are generally detected with only a few photons, mostly at energies close to the 50 GeV threshold. As such, a meaningful measurement of the spectral index is not possible. For cases in which the fit reaches the imposed upper bound (15), the photon index and its uncertainty are reported as null in the FITS file.}
\tablenotetext{b}{ When the flux is null, the flux uncertainty represents the 95\% Bayesian upper limit.}

\end{deluxetable}

%% file: tables/Table_new.tex
\begin{table*}[htbp]
\centering
\caption{4FHL sources not reported in previous {\it Fermi}-LAT catalogs. Galactic coordinates ($l$, $b$) and the positional uncertainty at 68\% C.L. ($R_{68}$) are in degrees. Observed photon index ($\Gamma$) and integrated flux above 50~GeV ($F_{50}$) include $1\sigma$ uncertainties. Note that the units of $F_{50}$ are $\times 10^{-12}$ ph cm$^{-2}$ s$^{-1}$. The closest 4FGL source column lists the nearest 4FGL-DR4 source and its angular separation (in degrees) from the 4FHL detection. The final column gives the 4FHL association and source class, when available.}
\begin{tabular}{lcccccccc}
\hline\hline
4FHL Name & $l$ & $b$ & $R_{68}$ & TS & $\Gamma$ & $F_{50}$ & Closest 4FGL Source & Assoc/Class \\
\hline
J0432.0+5603   & 150.34 &   5.41 & 0.02 & 28 & $1.74 \pm 0.44$ & $9.2 \pm 3.4$ & J0425.6+5522e (1.12$^\circ$)& 1LHAASO J0428+5331 (spp) \\
J0501.2+4432   & 162.15 &   1.51 & 0.03 & 33 & $1.89 \pm 0.48$ & $10.6 \pm 4.3$ & J0501.7+4459 (0.47$^\circ$)  & 1LHAASO J0500+4454 \\
J0810.9+3533   & 185.83 &  30.75 & 0.11 & 31 & $0.69 \pm 0.59$ & $6.1 \pm 3.4$ & J0809.6+3455 (0.68$^\circ$) & -- \\
J0823.3+1527   & 208.71 &  27.16 & 0.02 & 30 & $7.05 \pm 3.62$ & $4.7 \pm 2.7$ & J0829.0+1755 (2.83$^\circ$) & WISEA J082323.24+152447.9 (bll) \\
J1021.7$-$0225 & 246.28 &  43.38 & 0.02 & 29 & $4.18 \pm 1.61$ & $6.7 \pm 3.6$ & J1022.7$-$0112 (1.25$^\circ$) & -- \\
J1055.7$-$1231 & 263.91 &  41.41 & 0.03 & 26 & $2.10 \pm 0.72$ & $6.0 \pm 3.2$ & J1053.3$-$1134 (1.12$^\circ$) & -- \\
J1120.5$-$2648 & 278.99 &  31.80 & 0.02 & 52 & $3.50 \pm 0.93$ & $12.2 \pm 4.5$ & J1120.1$-$2645 (0.09$^\circ$) & NVSS J112031$-$264828 (bcu) \\
J1128.2$-$4919 & 289.27 &  11.33 & 0.02 & 29&  $2.98 \pm 1.02$ & $5.8 \pm 3.1$ & J1127.6$-$4920 (0.10$^\circ$)  & -- \\
J1213.2$-$6253 & 298.60 &  $-$0.35 & 0.03 & 31 & $1.95 \pm 0.62$ & $14.2 \pm 6.9$ & J1213.3$-$6240e (0.22$^\circ$) & SNR G298.5$-$0.3 (spp) \\
J1314.8+2358   &   3.34 &  83.85 & 0.03 & 28  & $12.09 \pm 6.45$& $4.9 \pm 2.7$ & J1314.7+2348 (0.16$^\circ$)  & -- \\
J1339.4+1150   & 341.67 &  71.00 & 0.03 & 25 & $2.90 \pm 0.99$ & $6.3 \pm 3.3$  & J1338.9+1153 (0.15$^\circ$) & -- \\
J1353.9+0152   & 336.24 &  60.64 & 0.03 & 27 & $4.38 \pm 1.92$ & $5.1 \pm 2.9$ & J1356.6+0234 (0.97$^\circ$) & PKS 1351+021 (fsrq) \\
J1450.1$-$1246 & 342.55 &  40.80 & 0.03 & 27 & $3.73 \pm 1.39$ & $7.4 \pm 3.9$  & J1453.0$-$1318 (0.90$^\circ$) & -- \\
J1503.7$-$4141 & 327.86 &  14.69 & 0.02 & 48 & $0.99 \pm 0.41$ & $11.1 \pm 4.6$ & J1503.6$-$4146 (0.10$^\circ$) & SN 1006 (snr) \\
J1605.2+5420   &  84.34 &  45.61 & 0.03 & 25 & $2.96 \pm 1.17$ & $3.1 \pm 1.6$ & J1605.5+5423 (0.07$^\circ$) & 3HSP J160519.1+542059 (bll) \\
J1718.4$-$2940 & 356.14 &   4.56 & 0.08 & 27 & $2.12 \pm 0.46$ & $18.1 \pm 6.6$ & J1719.2$-$2943 (0.19$^\circ$) & SNR G356.2+04.5 (snr) \\
J1729.9$-$3336 & 354.26 &   0.33 & 0.02 & 27 & $1.95 \pm 0.45$ & $19.4 \pm 7.0$ & J1730.5$-$3352 (0.31$^\circ$) &  NVSS J172951$-$333616 \\
J1749.4+4630   &  73.20 &  29.63 & 0.03 & 29 & $6.51 \pm 3.20$ & $3.4 \pm 2.0$ & J1750.2+4704 (0.58$^\circ$) & 3HSP J174929.9+463135 (bll) \\
J2024.6$-$0847 &  35.60 & $-24.69$& 0.02 & 30 & $3.96 \pm 1.57$ & $6.8 \pm 3.6$ & J2024.4$-$0847 (0.06$^\circ$) & 3HSP J202429.4$-$084804 (bll) \\
J2109.7$-$0253 &  47.50 & $-31.83$& 0.02 & 35 & $5.10 \pm 2.28$ & $7.6 \pm 3.7$ & J2108.7$-$0250 (0.26$^\circ$) & -- \\
\hline
\end{tabular}
\label{tab:4fhl_not_in_fermi}
\end{table*}